\documentclass[]{aa}

\usepackage{graphicx}
\usepackage{amsmath,amssymb,amsbsy}
\usepackage{natbib}
\usepackage{txfonts}
\usepackage{latexsym}
\usepackage{rotating}
\usepackage{booktabs,caption,fixltx2e}
\usepackage[flushleft]{threeparttable}
\usepackage{float}
\usepackage[toc,page]{appendix}
\usepackage{footnote}
\usepackage{color}

\newcommand{\Halpha}{H\ensuremath{\alpha}}
\newcommand{\Hbeta}{H\ensuremath{\beta}}
\newcommand{\Hgamma}{H\ensuremath{\gamma}}
\newcommand{\ionp}{\ion{P}{v}}
\newcommand{\ionsi}{\ion{Si}{iv}}

\newcommand{\ionheii}{\ion{He}{ii}}
\newcommand{\ionos}{\ion{O}{vi}}
\newcommand{\ionoc}{\ion{O}{iv}}

\newcommand{\mdot}{\ensuremath{\dot M}}

\newcommand{\vel}{\ensuremath{\varv}}
\newcommand{\vinfty}{\ensuremath{\vel_\infty}}
\newcommand{\veld}{\ensuremath{\vel_\text{D}}}

\newcommand{\velmin}{\ensuremath{\vel_\text{min}}}
\newcommand{\veldis}{\ensuremath{\vel_\text{dis}}}

\newcommand{\rmin}{\ensuremath{r_\text{min}}}
\newcommand{\rmax}{\ensuremath{r_\text{max}}}
\newcommand{\rcl}{\ensuremath{r_\text{cl}}}

\newcommand{\ncl}{\ensuremath{N_{\text{cl}}}}

\newcommand{\Teff}{\mbox{$T_\mathrm{eff}$}}

\newcommand{\kpc}{\ensuremath{\mathrm{kpc}}}
\newcommand{\R}{\ensuremath{R_{*}}}

\newcommand{\Rs}{\ensuremath{R_{\odot}}}
\newcommand{\Ls}{\ensuremath{L_{\odot}}}

\newcommand{\kms}{\ensuremath{\mathrm{km}/\mathrm{s}}}
\newcommand{\ms}{\ensuremath{\mathrm{M}_{\odot}}}
\newcommand{\msr}{\ensuremath{\ms/\mathrm{yr}}}

\defcitealias{Surlan:EtAl:2012}{Paper~I}

\allowdisplaybreaks

\newcommand{\Ondrejov}{Ond\v{r}ejov}
\newcommand{\dbetlaw}{double-\ensuremath{\beta}-law}

\newcommand{\POWR}{{\tt PoWR}}

\newcommand\backmatter{\appendix
\def\chaptermark##1{\markboth{%
\ifnum  \c@secnumdepth > \m@ne  \@chapapp\ \thechapter:  \fi  ##1}{%
\ifnum  \c@secnumdepth > \m@ne  \@chapapp\ \thechapter:  \fi  ##1}}%
\def\sectionmark##1{\relax}}

\begin{document}

\title{Macroclumping as solution of the discrepancy between {\Halpha} and {\ionp}
mass loss diagnostics for O-type stars\thanks{Based on observations collected with the 
Perek 2-m Telescope of the Ondřejov Observatory, Czech Republic.}\thanks{Based on observations
taken at Complejo Astron\'omico El Leoncito (CASLEO), operated under an
agreement between the Consejo Nacional de Investigaciones Cient\'{\i}ficas y
T\'ecnicas de la Rep\'ublica Argentina, the Secretar\'{\i}a de Ciencia y
Tecnolog\'{\i}a de la Naci\'on and the National Universities of La Plata,
C\'ordoba and San Juan.}}
\author{}
\author{B. \v{S}urlan\inst{1,2}
\and W.-R. Hamann\inst{3}
\and A. Aret\inst{4}
\and J. Kub\'at\inst{1}
\and L. M. Oskinova\inst{3}
\and A. F. Torres\inst{5,6}
}

\institute{Astronomick\'y \'ustav, Akademie v\v{e}d \v{C}esk\'e
	Republiky, CZ-251 65 Ond\v{r}ejov, Czech Republic
\and	Matemati\v{c}ki Institut SANU, Kneza Mihaila 36, 11001 Beograd,
	Republic of Serbia
\and
	Institut f\"ur Physik und Astronomie, Universit\"at Potsdam,
	Karl-Liebknecht-Stra{\ss}e 24/25, 14476 Potsdam-Golm, Germany
\and    Tartu Observatory, 61602, T\~{o}ravere, Tartumaa, Estonia
\and    Departamento de Espectroscop\'ia, Facultad de Ciencias Astron\'omicas 
        y Geof\'isicas, Universidad Nacional de La Plata, Paseo del Bosque S/N, 
        La Plata, B1900FWA, Buenos Aires, Argentina
\and    Instituto de Astrof\'isica de La Plata (CCT La Plata - CONICET,
        UNLP), Paseo del Bosque S/N, La Plata, B1900FWA, Buenos Aires, Argentina 
}

\abstract
%
%
{Recent studies of O-type stars demonstrated that discrepant mass-loss rates
are obtained when different diagnostic methods are  employed -- fitting
the unsaturated UV resonance lines (e.g. {\ionp}) gives drastically
lower values than obtained from the {\Halpha} emission. Wind inhomogeneity 
(so-called ``clumping'') may be the main cause for this discrepancy. } 
%
%
{In a previous paper, we have presented \mbox{3-D} Monte-Carlo calculations for the
formation of scattering lines in a clumped stellar wind. In the present
paper we select five O-type supergiants (from O4 to O7) and test
whether the reported discrepancies can be resolved this way.} 
%
%
{In the first step, the analyses start with simulating the observed
spectra with Potsdam Wolf-Rayet  ({\POWR}) non-LTE model
atmospheres. The mass-loss rates are adjusted  to fit best to the
observed {\Halpha} emission lines.  For the unsaturated UV resonance
lines (i.e. {\ionp}) we then apply our \mbox{3-D} Monte-Carlo code, which
can account for wind clumps of any optical depths (``macroclumping''), a
non-void inter-clump medium, and a velocity dispersion inside the 
clumps. The ionization stratifications and underlying photospheric
spectra are adopted from the {\POWR} models. From fitting the observed
resonance line profiles, the properties of the wind clumps
are constrained.}
%
%
{Our results show that with the mass-loss rates that fit {\Halpha} (and
other Balmer and {\ionheii} lines), the UV resonance lines (especially
the unsaturated doublet of {\ionp}) can also be reproduced
without problem when macroclumping is taken into account. There is no
need to artificially reduce the mass-loss rates, nor to assume a
sub-solar phosphorus abundance or an extremely high clumping factor, 
contrary to what was claimed by other authors. These consistent mass-loss 
rates are lower by a factor of 1.3 to 2.6, compared to the mass-loss
rate recipe from  Vink et al.}
%
%
{Macroclumping resolves the previously reported discrepancy between
{\Halpha} and {\ionp} mass-loss diagnostics.}
  

\maketitle


\section{Introduction}

The most important property of massive, hot stars is their mass loss
expelled via stellar winds. These winds can be extremely strong, which
imposes a significant effect on their evolution, and affects their
surface abundances \citep[for a review see, e.g.,][and references therein]
{Meynet:Maeder:2007}.

The line-driven wind theory, first proposed by \cite{LucySolomon:1970}
and later elaborated  by \citet*{Castor:EtAl:1975}, can explain the
physical mechanism by which massive stars lose mass. However, the
accurate values of the wind parameters are still under debate. One of
the most challenging problems is the determination of reliable  
mass-loss rates. The latter are derived from observations with the aid
of some physical models. However, discordances  between different
mass-loss rate diagnostics were found  \citep[for a review
see][]{Puls:EtAl:2008}. 

A major complication of mass-loss estimates comes from the fact that 
stellar winds are inhomogeneous, as indicated by several
observational evidences \citep[see, e.g.,][]{Eversberg:EtAl:1998,
Lepine:Moffat:2008, Prinja:Massa:2010}, but also predicted by numerical simulations
\citep[e.g.,][]{Feldmeier:EtAl:1997, Runacres:Owocki:2002, 
DessartOwocki:2005}. These simulations show that the instability of
wind line driving leads to the formation of shocks and spatial
structures in both density and velocity, i.e.\ clumps. 

Analyses of massive star spectra rely on their comparison with
sophisticated model simulations. State-of-the-art model atmosphere
codes account for non-LTE radiative transfer in a spherically symmetric
wind, and incorporate detailed model atoms plus an approximate
treatment of the line-blanketing effect from iron-group elements. A few
such codes have been developed, e.g.\ {\tt CMFGEN}
\citep{HillierMiller:1998}, {\POWR} \citep{Hamann:Grafener:2004}, and
{\tt FASTWIND} \citep{Puls:EtAl:2005}. In all these codes, clumping is
included only in the approximation that the clumps are assumed to be
optically  thin at all frequencies (the so called ``microclumping''
approximation). The clumps have a density that is enhanced by a
``clumping factor'' $D$ compared to a smooth wind with the same mass-loss
rate. The clumps move according to the adopted velocity law of the
wind. In most simulations, the inter-clump space is assumed to be void.

The main effect of microclumping is that empirical mass-loss rates
which are derived from recombination lines, i.e. from a process that 
depends {\em quadratically} on density, are overestimated by a factor
of $\sqrt{D}$ when microclumping is neglected, and have to be reduced
accordingly by a mild factor of about 2 or 3. However,
\cite{Massa:EtAl:2003} and \cite{Fullerton:EtAl:2006} studied O-star
spectra in the far-ultraviolet (FUV) obtained with the  Far Ultraviolet
Spectroscopic Explorer ({\tt FUSE}) satellite, which show the
unsaturated resonance line doublet of {\ionp} at $\lambda\lambda~1118,
1128~$\AA. The formation of resonance lines depends only linearly on
density, and is therefore not sensitive to microclumping. They found 
mass-loss rates that are lower by factors 10 to 100, compared with those
obtained from recombination lines under the assumption of no clumping. 
Consequently, these authors conclude that the mass-loss rates have
been greatly over-estimated when derived from recombination lines,
obviously because the clumping contrast $D$ is in fact extremely high
(and the volume filling factor of the clumps correspondingly
tiny). Such low mass-loss rates would have far-reaching consequences,
e.g. for the evolution of massive stars.  

The same strategy to resolve the mass-loss rate discrepancy was
employed  in papers by \cite{Bouret:EtAl:2003, Bouret:EtAl:2005}. They
also reduced the mass-loss rates in order to weaken the UV resonance
lines, while the {\Halpha} emission is kept at the observed
strength by adopting extremely large clumping factors up to $D=100$.

As an additional means to achieve a consistent fit,
\cite{Bouret:EtAl:2012} reduced the phosphorus abundances to values that
are lower by a factor 1.4 to 5.1 than the solar abundance according to
\cite{Asplund:EtAl:2005}. However, sub-solar abundances for O stars are 
not expected and have no justification.

Alternatively, \cite{Oskinova:EtAl:2007} suggested that the mass-loss
rate discrepancy can be explained by the effect of optically thick
clumps, which was hitherto neglected within the microclumping
approximation.  They promoted the ``macroclumping'' (porosity)
approach, taking into account that clumps may become optically thick 
at certain frequencies. These authors showed that,
while the optically thin {\Halpha} line is not affected by wind
porosity, the {\ionp} resonance doublet becomes significantly weaker
when macroclumping is included. This leads to the conclusion that 
clumping must be included in the modeling to get different diagnostics 
to agree on the mass-loss rate and that
the microclumping approximation is not adequate for modeling 
optically thick transitions.
 
To derive more reliable mass-loss rates from observation, and to
resolve discrepancy between different diagnostics, a more detailed
treatment of wind  clumping is needed. Unfortunately, a general
treatment of \mbox{3-D} clumps in full non-LTE radiative transfer
simulations is not possible. However, the formation of resonance lines
can be treated in the much simpler pure-scattering approximation. This
allows the use of Monte-Carlo techniques, which can be adapted to
complicate geometrical situations.  
 
\cite{Sundqvist:EtAl:2010, Sundqvist:EtAl:2011} used a \mbox{2-D}
and pseudo-\mbox{3-D} stochastic wind model, and achieved a reasonable
line fit for HD\,210839 which is also in our sample (see below).

\cite{Surlan2:EtAl:2012, Surlan:EtAl:2012} developed a full \mbox{3-D}
description of clumping, and investigated how the properties of clumping
(e.g., the velocity dispersion inside the clumps, the radius where clumping
sets on, and the density of the inter-clump medium) affect the resonance
line profiles and, consequently, the derived mass-loss rates.

The intention of the present paper is to check for a small sample of
stars whether our detailed treatment of clumping, together with solar 
phosphorus abundance and moderate $D$, may resolve the
discordance between the mass-loss rates derived from  {\Halpha} and
{\ionp} diagnostics, and also to establish some global properties of
wind  clumping. We selected 5 O-type supergiants and analyzed their
spectra first by means of the Potsdam Wolf-Rayet ({\POWR}) model
atmosphere code, and then applied our \mbox{3-D} Monte-Carlo radiative 
transfer code for simulating the UV resonance lines. 

In the following section we present our stellar sample, observations,
and data  reduction. The {\POWR} models and the \mbox{3-D} Monte-Carlo
code for the resonance line formation in a clumped wind are introduced
in  Sect.~\ref{powrmodel} and Sect.~\ref{tridmodel}, respectively. Our
spectral fitting procedure is explained in Sect.~\ref{modcalc}.  The
results of the  consistent analysis of the optical and UV spectra are
presented in Sect.~\ref{results} and  discussed in Sect.~\ref{diskuse}.
Finally, a summary is given in Sect.~\ref{conclusion}. The complete
spectral  fits are available in the {\em Online Material}.

\section{Stellar sample and observation}

\subsection{Stellar sample}

\begin{table*}[t]
\centering
\caption {Input stellar and wind parameters and element abundances by mass fraction.} 
\label{tab:inputstelwindpara} 
\small{\begin{tabular}{l c c c c c c c c | c c c c c} 	
\hline
\hline
\rule[0mm]{0mm}{3.5mm} \!\!\!
Star        & Spec.   & $\Teff$    & $\log{g}$ & $\R$   & $\log{\frac{L}{\Ls}}$     & $\beta_{1}$     & $\vinfty$& Ref. &   H    &  He   &  C  &  N &  O   \\
                      &  type      & [kK]      &        & [$\Rs$]  &                           &     &  [$\kms$]       &    &        &       &     &    &      \\         
\hline			 
\hline
\rule[0mm]{0mm}{3.5mm} \!\!\!
HD\,66811  & O4I(n)f        & 39.0             & 3.55          &  19.6    & 5.90                      & 0.70          &  2250  &  1 & 0.61   & 0.37  &  2.86E-03   & 1.05E-02 & 1.30E-03  \\ 
HD\,15570  & O4If           & 38.0             & 3.28          &  21.6    & 5.94                      & 1.10          &  2200  &  2 & 0.71   & 0.28  &  3.27E-03   & 4.79E-03 & 2.63E-03  \\
HD\,14947  & O4.5If            & 37.5             & 3.45          &  26.6    & 6.09                      & 0.95          &  2350  &  3 & 0.68   & 0.31  &  1.66E-03   & 5.00E-03 & 1.44E-03  \\ 
HD\,210839 & O6.5I(n)fp        & 36.0             & 3.55          &  23.3    & 5.91                      & 1.00          &  2250  &  3 & 0.68   & 0.31  &  1.32E-03   & 4.67E-03 & 3.23E-03  \\ 
HD\,192639 & O7.5Iabf         & 35.0             & 3.45          &  18.5    & 5.66                      & 0.90          &  2150  &  3 & 0.62   & 0.37  &  1.09E-03   & 5.01E-03 & 4.01E-03  \\ 
\hline
\end{tabular}}
\tablefoot{{Gravitational accelerations, $\log{g}$, are effective values.}
\tablebib{Stellar and wind parameters are taken from: (1)~\citet{Oskinova:EtAl:2007}; (2)~\citet{Bouret:EtAl:2012}; (3)~\citet{Puls:EtAl:2006}. Spectral type of HD\,66811 is taken from \cite{Walborn:EtAl:2009}. For other stars spectral types are taken from \cite{Sota:EtAl:2011}.}}
\end{table*}

We selected 5 O-type Galactic supergiants covering spectral types O4If
to O7If (see Table~\ref{tab:inputstelwindpara}). All  these stars are
very luminous and show evidence of a strong wind. Due to intrinsic 
instability of the line driving mechanism it is expected that these
winds should exhibit pronounced clumping.

These stars have been analyzed already in the optical, UV, infrared and
radio spectral regions \citep[e.g.,][]{Markova:EtAl:2004,
Repolust:EtAl:2005, Puls:EtAl:2006, Bouret:EtAl:2012}. Stellar
parameters of  the sample were reliably determined, and also mass-loss
rates from  different diagnostics are available for comparison.  All
stars from our sample are presumably single, showing no indications of
binarity  \citep{Bouret:EtAl:2012, Mason:EtAl:1998, DeBecker:EtAl:2009,
Turner:EtAl:2008}. For all stars of the sample, FUV spectra that
include the {\ionp} resonance doublet are available, which is
prerequisite to study the so-called ``{\ionp} problem''.

\subsection{Optical spectra}

\begin{table*}
\centering
\caption {Optical, FUV, and NUV observation 
logs.} 
\label{tab:obslog} 
\small{\begin{tabular}{|c | c c c c | c c | c c |} 
\hline
\hline
\rule[0mm]{0mm}{3.5mm}
Star     & \multicolumn{4}{|c|}{Optical} &  \multicolumn{2}{c}{FUV}  &  \multicolumn{2}{c|}{NUV}  \\
\hline
\rule[0mm]{0mm}{3.5mm}
HD  &  UT date  & UT start & $t_{\text{exp}}$ [s] & wavelength [{\AA}] & Data set & UT date  &  Date ID   & UT data \\
\hline
\hline
\rule[0mm]{0mm}{3.5mm}
  66811  & 2012-11-29   & 08:53:46   & 40     &    3850--7100        & C044-001 & 1973-02-22 & LWP13207HL  & 1988-11-05  \\
\hline
\rule[0mm]{0mm}{3.5mm}
  15570  &  2012-12-30  & 20:27:54 & 2700 & 6254--6764 & E0820101 & 2005-11-08 & LWR04941LL  & 1979-07-04  \\
         &  2013-08-06  & 00:37:59 & 3600 & 4656--4907   &          &            & SWP04112LL  & 1979-02-01  \\
%
\hline\rule[0mm]{0mm}{3.5mm}
  14947  &  2012-12-29  & 22:58:10 & 3600 & 6254--6764  & E0820201 & 2004-09-30 & LWR07220LL  & 1980-03-17  \\  
         &  2013-08-05  & 23:33:22 & 3600 & 4656--4907   &          &            & SWP02876LL  & 1978-10-07  \\ 
%
\hline\rule[0mm]{0mm}{3.5mm}
 210839  &  2013-01-12  & 20:29:47 & 1800 & 6254--6764  & P1163101 & 2000-07-22 & LWR15139LL  & 1983-01-28  \\ 
         &  2013-01-12  & 18:10:58 & 3600 & 4656--4907  &          &            & SPW04015HL  & 1979-01-24  \\   
%
\hline\rule[0mm]{0mm}{3.5mm}
 192639  &  2012-12-03  & 18:05:26 & 3600 & 6254--6764  & P1162401 & 2000-06-12 & LWP03192LL  & 1984-04-21  \\ 
         &  2012-12-30  & 17:13:13 & 3600 & 4656--4907  &          &            & SWP22808LL  & 1984-04-21  \\
\hline
\end{tabular}}
\end{table*}

For the four northern stars of our sample, {\Halpha} and blue
spectra were observed with a CCD SITe ST-005 800$\times$2000 pix
camera attached to the Coud\'e spectrograph of the \mbox{2-m} 
telescope at the {\Ondrejov} Observatory (Czech Republic), with the
slit width set to $0.6{\arcsec}$. 
Two grating angles were chosen, one
centered at {\Halpha} line and the second one covering {\ionheii} 4686\,\AA\
and {\Hbeta} lines. The achieved spectral resolution is 13\,600 
and 19\,400, respectively.

All spectra were wavelength calibrated with a ThAr comparison arc
spectra obtained shortly after each exposure. The telluric features in
the spectra were removed using
spectra of
the fast-rotating stars 27 Vul and 116 Tau.
The data  reduction (including telluric and heliocentric velocity
corrections) was performed with standard IRAF\footnote{IRAF is distributed
by the National Optical Astronomy Observatories, which are operated by
the Association of Universities for Research in  Astronomy, Inc., under
cooperative agreement with the National Science Foundation.} tasks.
The program Cosmic Ray Removal\footnote{\tt
http://users.camk.edu.pl/pych/DCR/} ({\tt dcr}, \citealp{Pych:2004})
was used to clean the spectra.


The optical spectrum of HD\,66811 was taken at the Complejo Astron\'omico El
Leoncito (CASLEO) in Argentina. The observation was carried out with the
2.15-m Jorge Sahade telescope using a REOSC echelle spectrograph in
cross-dispersion mode with a Tek 1024 x 1024 pixel CCD as detector. The
adopted instrumental configuration was a 316 l/mm grating at a tilt angle of 
5$\,^{\circ}$40$\,^{\prime}$ and a 350 $\mu$m slit width, resulting in a resolving power of
12\,500. A ThAr lamp was used as a comparison source, with a reference
exposure taken immediately after the stellar target at the same telescope
position. The data reduction was performed with standard IRAF tasks.
The spectroscopic observation logs are documented in Table~\ref{tab:obslog}.

\subsection{Ultraviolet spectra}

The spectral region of \ionp\ resonance doublet was covered
by high-resolution observations with FUSE, which we 
retrieved from the {\tt MAST}\footnote{\tt http://archive.stsci.edu}
archive  (see Table~\ref{tab:obslog}). 
To increase the S/N ratio, the {\ionp} spectra of HD\,15570 and 
HD\,14947 were smoothed using the splot task in IRAF. 
Low-resolution near-ultraviolet (NUV) spectra (1200 to 2000 {\AA}),
taken with the International Ultraviolet Explorer (IUE), were 
downloaded from the {\tt INES Archive Data Server}\footnote{\tt
http://sdc.cab.inta-csic.es}  (see Table~\ref{tab:obslog}). 
The FUV spectrum of HD\,66811 had been observed with the Copernicus
satellite. 

All observed spectra were corrected for the radial velocity of the
individual star before the comparison with model simulations.

\section{1-D spherically symmetric wind models}
\label{powrmodel}

To analyze the observed spectra  we calculated wind models using the
{\POWR} unified model atmospheres code \citep[see][and references
therein]{Hamann:Grafener:2004}. The {\POWR} code is able to solve
non-LTE radiative transfer in a spherically expanding atmosphere
simultaneously with the statistical equilibrium equations, and
accounts for energy conservation. Detailed model atoms of the most
relevant elements (H, He, C, N, O, Si, and P) are taken into account
in the present paper. Line blanketing is taken into account, with the 
iron-group elements being treated in the super-level approach. 
Mass-loss  rate and wind velocity are among the free parameters of 
the models. 

\subsection{Stellar parameters and chemical composition}

Stellar and wind parameters of the stars as obtained by
\cite{Puls:EtAl:2006}, \cite{Oskinova:EtAl:2007}, and
\cite{Bouret:EtAl:2012} serve as input parameters for our {\POWR}
models (see Table~\ref{tab:inputstelwindpara}). 
The chemical abundances of H, He, C, N, and O are adopted from
\cite{Bouret:EtAl:2012} (see Table~\ref{tab:inputstelwindpara}).  For
the mass fractions of Si, P, and Fe-group elements we take the solar
values  ($6.649\times 10^{-4}$, $5.825\times 10^{-3}$, and $1.292\times
10^{-3}$, respectively) as determined by \cite{Asplund:EtAl:2009}.

\subsection{Velocity field}
\label{sect:velfield}

\begin{figure}[t]
\includegraphics[width=0.49\textwidth]{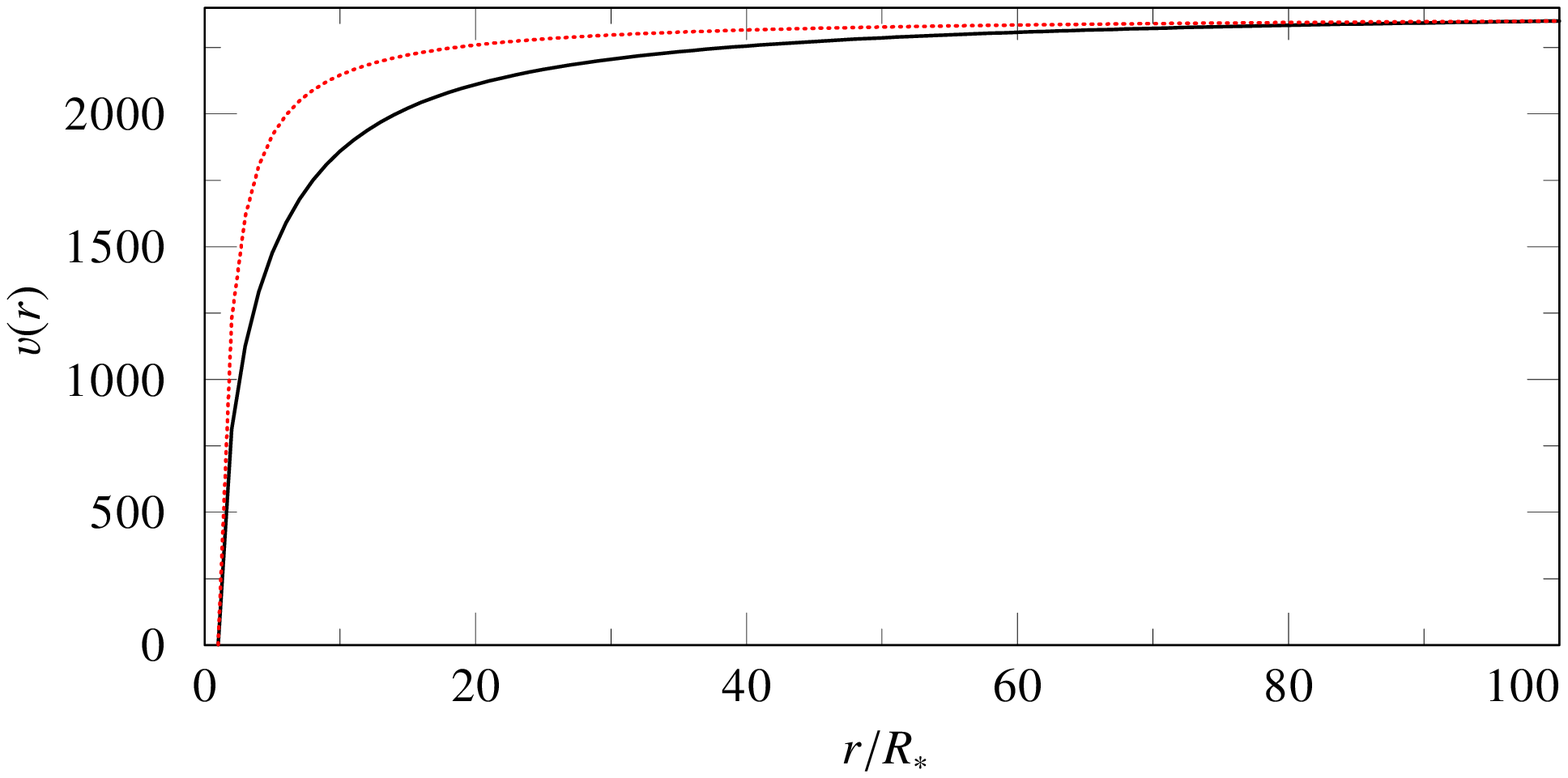}
\put(-120,50){\includegraphics[scale=0.2]{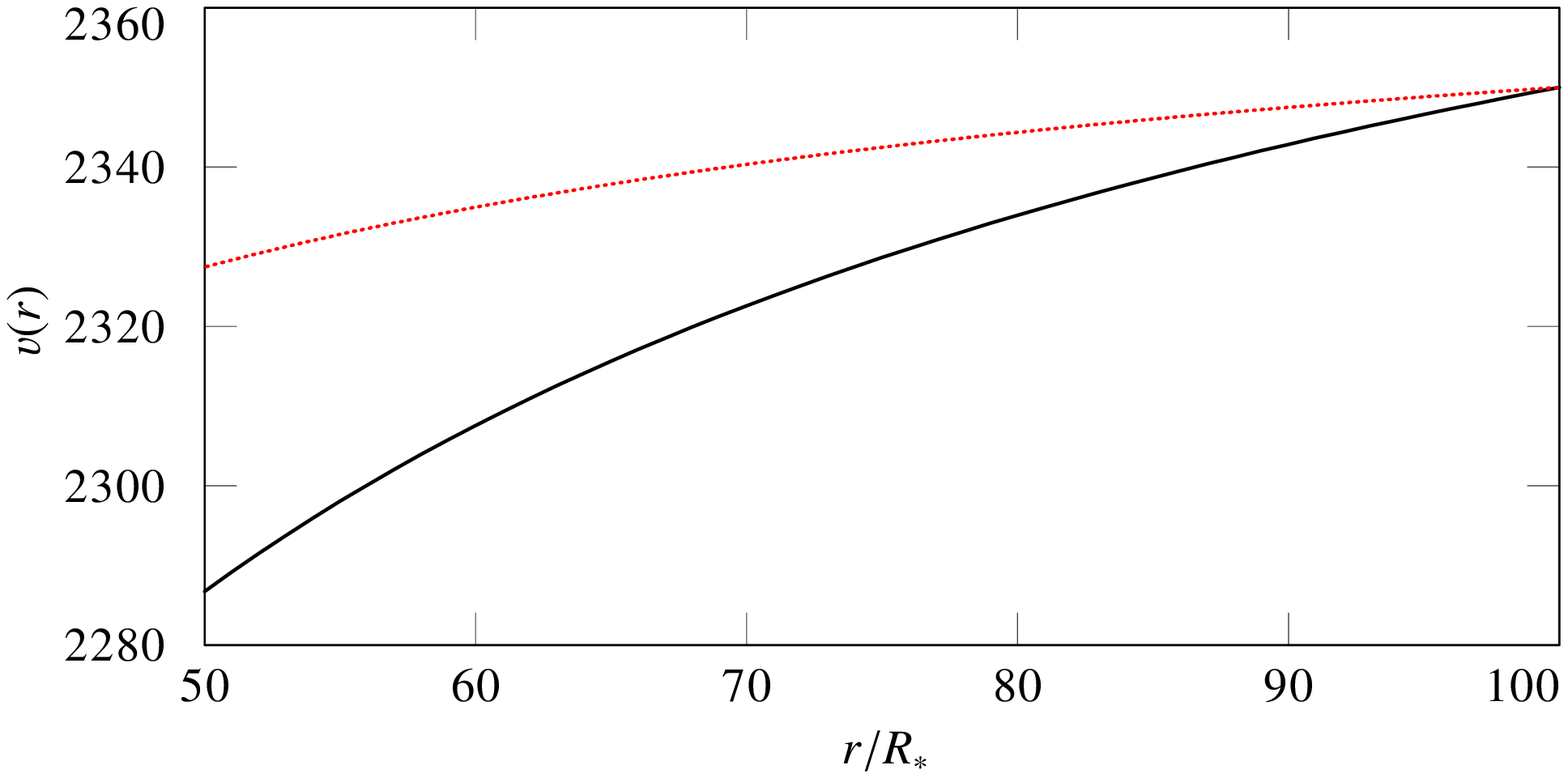}}\\
\caption{Dependence of wind velocity on radius.
Comparison of the standard (single) $\beta$-law (dotted-red line) and
{\dbetlaw} (solid-black line).}
\label{beta-ionfrac}
\end{figure}

The adopted velocity field in the model consists of two parts. In the
inner part, the  hydrostatic equation is integrated with the 
stratification of temperature and mean particle mass, yielding the
density stratification.  The velocity in this part of the wind is then
defined via the continuity equation. This hydrostatic part  of the
atmosphere is connected smoothly to the wind, with the so called
{\dbetlaw} \citep[][]{Hillier:Miller:1999}. 
The {\dbetlaw} consists of the sum of two beta-law terms 
with different exponents $\beta_{1}$ and $\beta_{2}$, each of them
contributing a pre-specified fraction to the total wind velocity. 
Compared to the standard ``one-beta'' law, this allows for 
a smaller velocity gradient in the lower part of the wind, while the
second term, for which we adopt always $\beta_{2}=6$ and a 
contribution of 35\% to the final velocity, causes some noticeable
acceleration even at relatively large distances from the star 
(Fig.~\ref{beta-ionfrac}). The values for {\vinfty} and $\beta_{1}$,
as included in Table~\ref{tab:inputstelwindpara}, were also adopted
from the references.

In the {\POWR} models the lines are broadened by thermal and
microturbulent motion with  $\varv_{\rm D} = 20\,\kms$.  In addition,
radiation  damping and pressure broadening are accounted for in the
formal integral.

\subsection{Clumping in the 1-D wind model}
\label{powrclump}

In our \POWR\ models, the wind inhomogeneities are treated in the
microclumping approximation  \citep[for more details
see][]{Hamann:Koesterke:1998}. The matter density in the clumps is
enhanced by a factor $D=1/f_{V}$, where $f_{V}$ is  the fraction of
volume filled by clumps. In the present study, we allow the clumping
factor to depend on radius, starting to deviate from the homogeneous
wind ($D=1$) at about the sonic point ($5~\kms$) 
and quickly reaching $D=10$ at $\varv(r) = 40~\kms$.

\section{3-D clumped wind model}
\label{tridmodel}

\begin{figure}[t]
\begin{center}
\includegraphics[width=0.9\columnwidth]{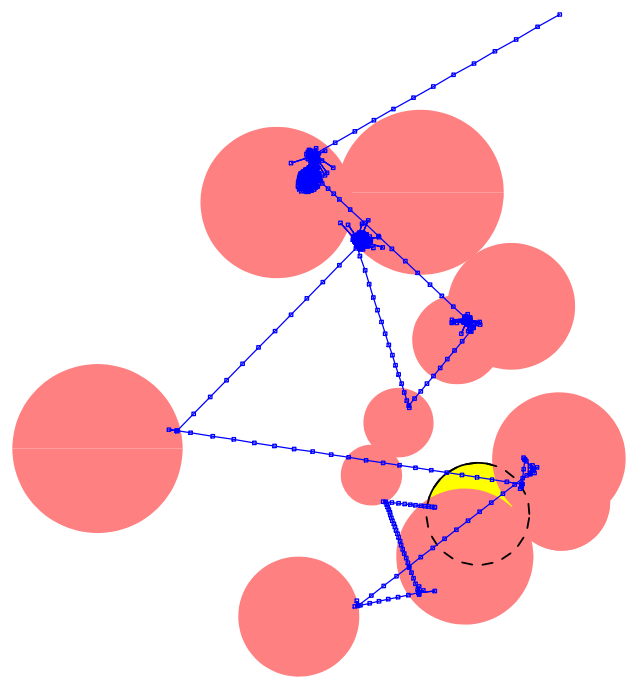}
\caption{Snapshot from a \mbox{3-D} Monte-Carlo simulation,
showing in two-dimensional projection an example of a clumps realization 
(red full circles) in respect to star (yellow dashed circle) and the path 
of a line-scattered photon (blue dotted line).}
\label{photon}
\end{center}
\end{figure}

To model the {\ionp} resonance line profiles, we use our Monte-Carlo code for
the radiative transfer in a three-dimensional clumped wind.
In the spirit of a core-halo approximation, the lower boundary of
the wind is placed just above the photosphere. Here we employ the
photospheric line spectrum as obtained from the {\POWR} model as inner
boundary. 

In the wind, we create a random distribution of spherical clumps. These
clumps move with the wind velocity law, but have also an additional
internal velocity gradient (see below). The number density of clumps
obeys the continuity equation. The density in the clumps and in the
inter-clump medium is specified from the mass-loss rate and the clumping
parameters. 

\begin{table}[t] 
\centering
\caption {Finally adjusted stellar and wind parameters.
$\mdot_{\text{vink}}/\mdot$ is the ratio of theoretical ($\mdot_{\text{vink}}$, \citealt{Vink:EtAl:2000}) to 
measured ($\mdot$) mass-loss rates.}  
\label{tab:derivedstelwindpara} 
\small{\begin{tabular}{l c c c c} 	
\hline
\hline
\rule[0mm]{0mm}{3.5mm} \!\!\!
Star                    & distance                   &  $R_{V}$        & $\mdot$             & $\mdot_{\text{vink}}/\mdot$ \\ 
                        & [$\kpc$]                   &                 & [$10^{-6}~\msr$]    &                             \\  
\hline
\hline
\rule[0mm]{0mm}{3.5mm} \!\!\!
HD 66811                &  2.34     &  3.10          & 2.51            &   1.86                     \\  
HD 15570                &  2.34     &  3.10          & 2.75            &   2.58                     \\  
HD 14947                &  3.00     &  2.80          & 2.82            &   2.32                     \\  
HD 210839               &  0.95     &  3.15          & 1.62            &   1.78                     \\  
HD 192639               &  2.00     &  3.10          & 1.26            &   1.34                     \\  
\hline
\end{tabular}}
\end{table}

The photons which are now released at the lower boundary travel through
the wind, where they can be repeatedly scattered in the considered line
doublet (continuum opacities are neglected). The line scattering is
assumed to be isotropic in the co-moving frame of reference, while the
frequencies are completely redistributed over the Doppler-broadened
opacity profile. The opacity is computed according to the mass-loss
rate, element abundance, and ionization fraction. The latter is
retrieved from the {\POWR} model. Traveling photons experience Doppler
shifts due to the wind expansion. Once a photon crosses the outer
boundary of the wind it is counted for the emergent profile. The
principle of this formalism is illustrated in Fig.~\ref{photon}, while
more details of the code are given in \cite{Surlan:EtAl:2012}.

\begin{table}[t]
\caption{Fixed model parameters used in the \mbox{3-D} Monte-Carlo code.} 
\label{tab:fixclumpara}
\begin{center}
\begin{tabular}{ l  rcl }
\hline  
\hline 
\rule[0mm]{0mm}{3.5mm} \!\!\!
Model parameters  &   \multicolumn {3}{c}{Value~~~~~~~~~~~} \\
\hline
\hline
\rule[0mm]{0mm}{3.5mm} \!\!\!
Inner boundary of the wind       &$\rmin$&=&$1\,\R$ \\
Outer boundary of the wind       &$\rmax$&=&$100\,\R$    \\
Clump separation parameter       &$L_{0}$&=&0.5      \\
Clumping factor                  &$D$&=&$10$           \\
Onset of clumping                &$\rcl$&=&$1\,\R$         \\
Velocity at the photosphere      &$\velmin$&=&10\,[\kms]\\
Doppler velocity                 &$\veld$&=&20 [\kms] \\              
\hline
\end{tabular}
\end{center}
\end{table}

\begin{figure*}[!t]
\includegraphics[width=0.5\textwidth]{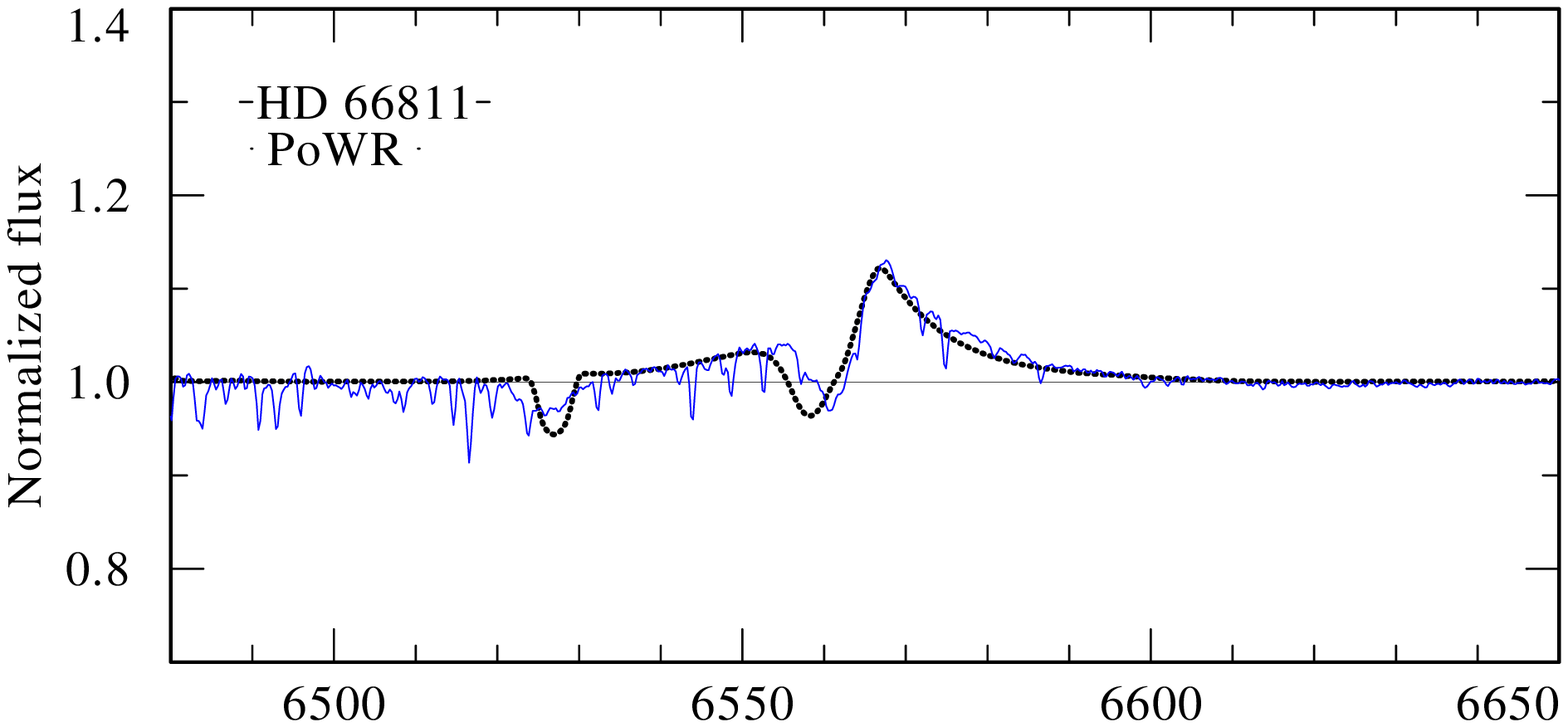}
\includegraphics[width=0.5\textwidth]{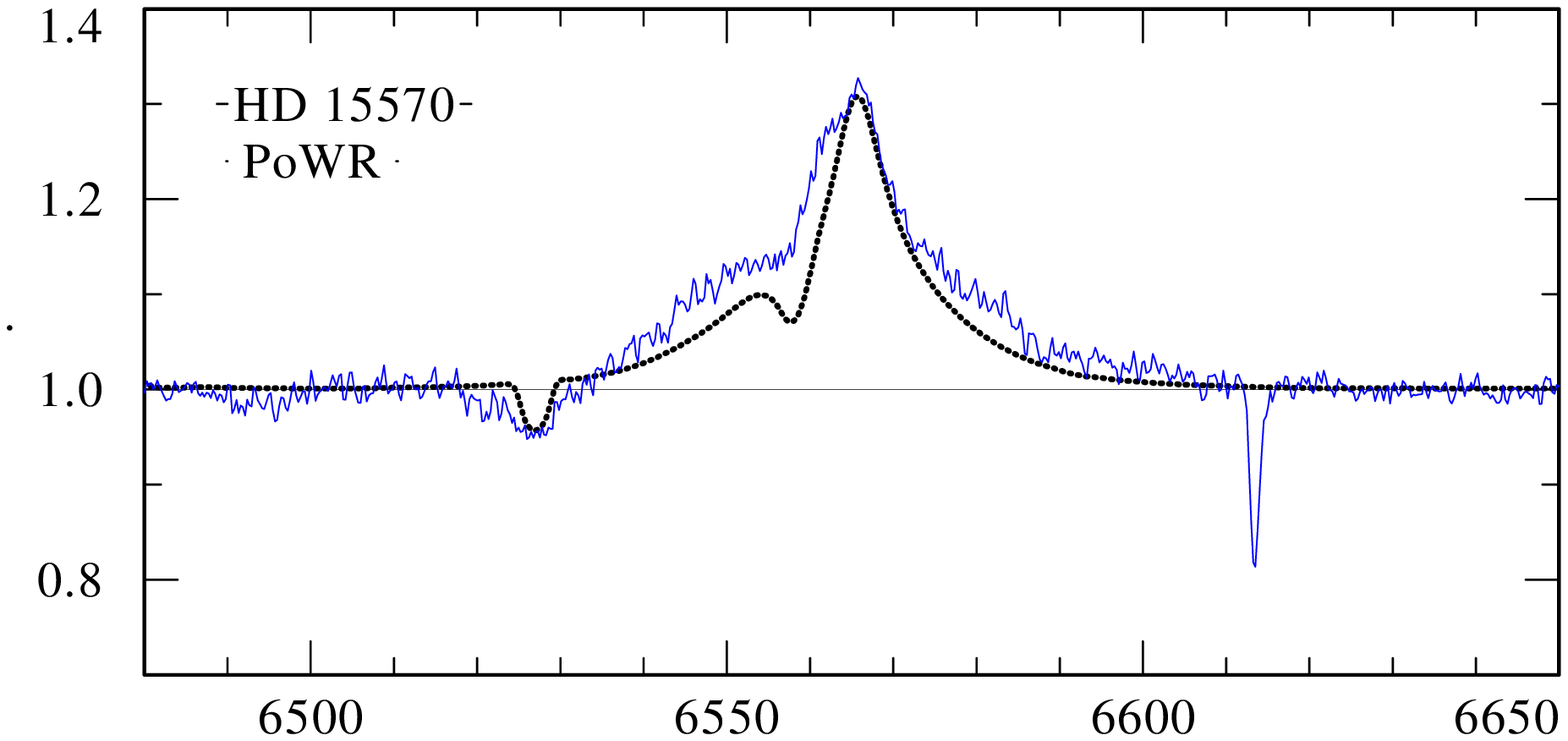}
\includegraphics[width=0.5\textwidth]{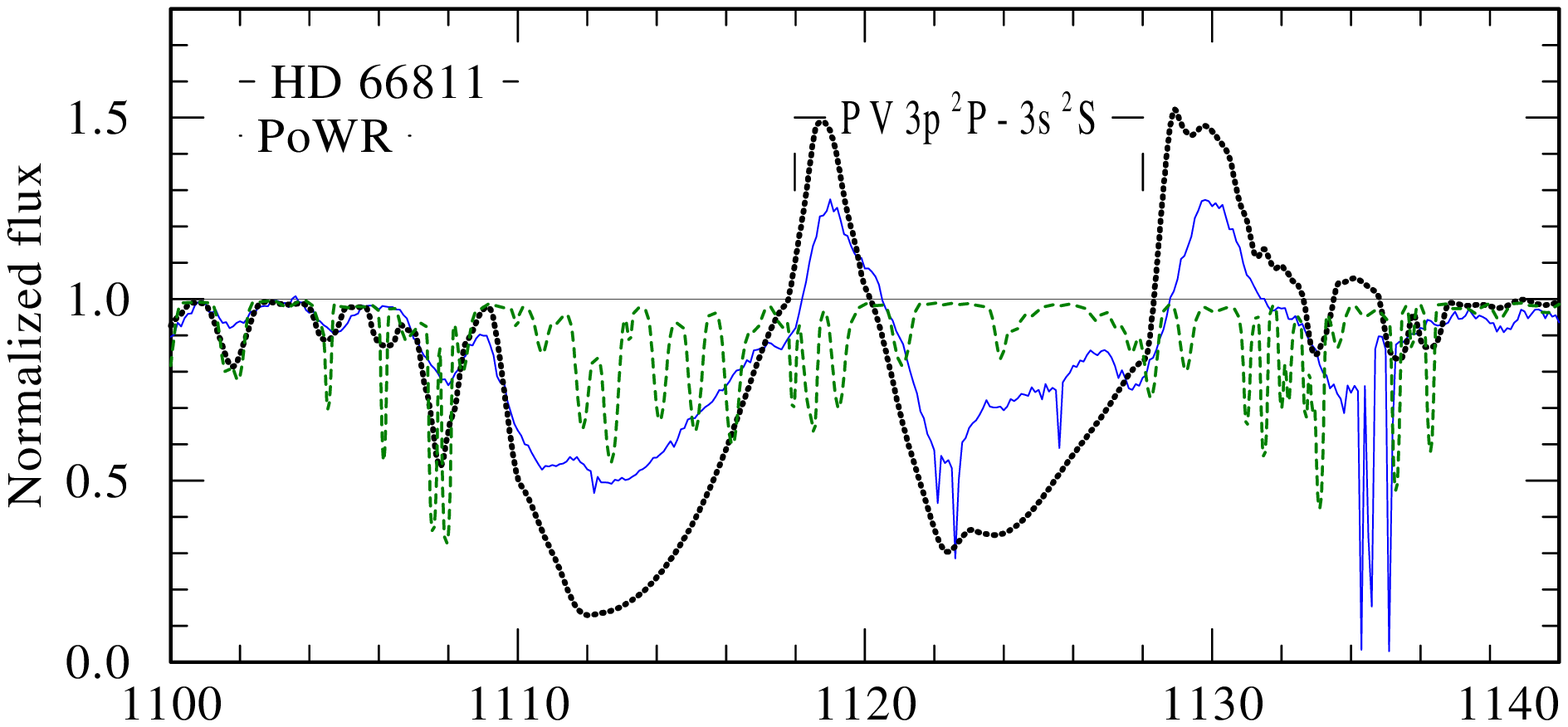}
\includegraphics[width=0.5\textwidth]{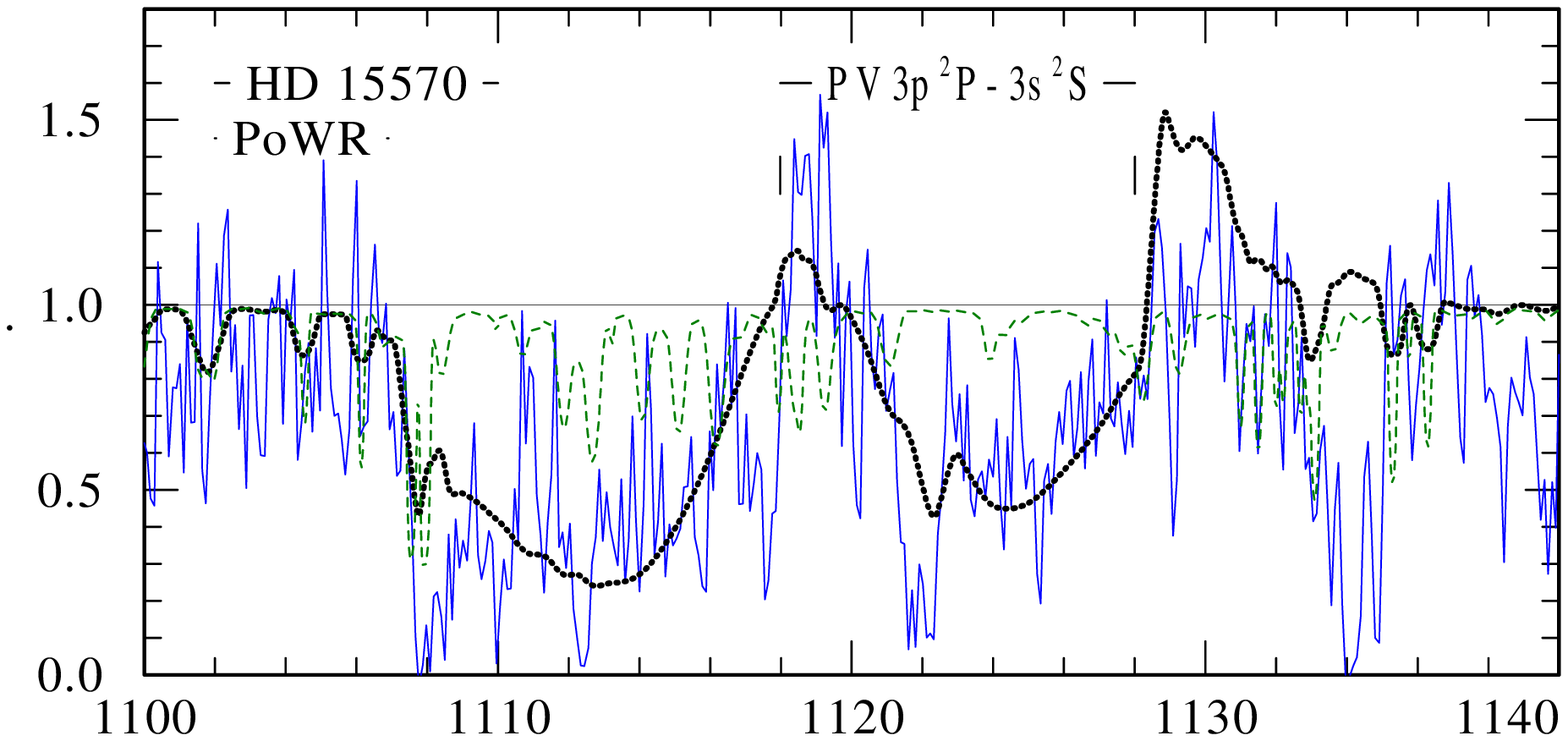}
\includegraphics[width=0.5\textwidth]{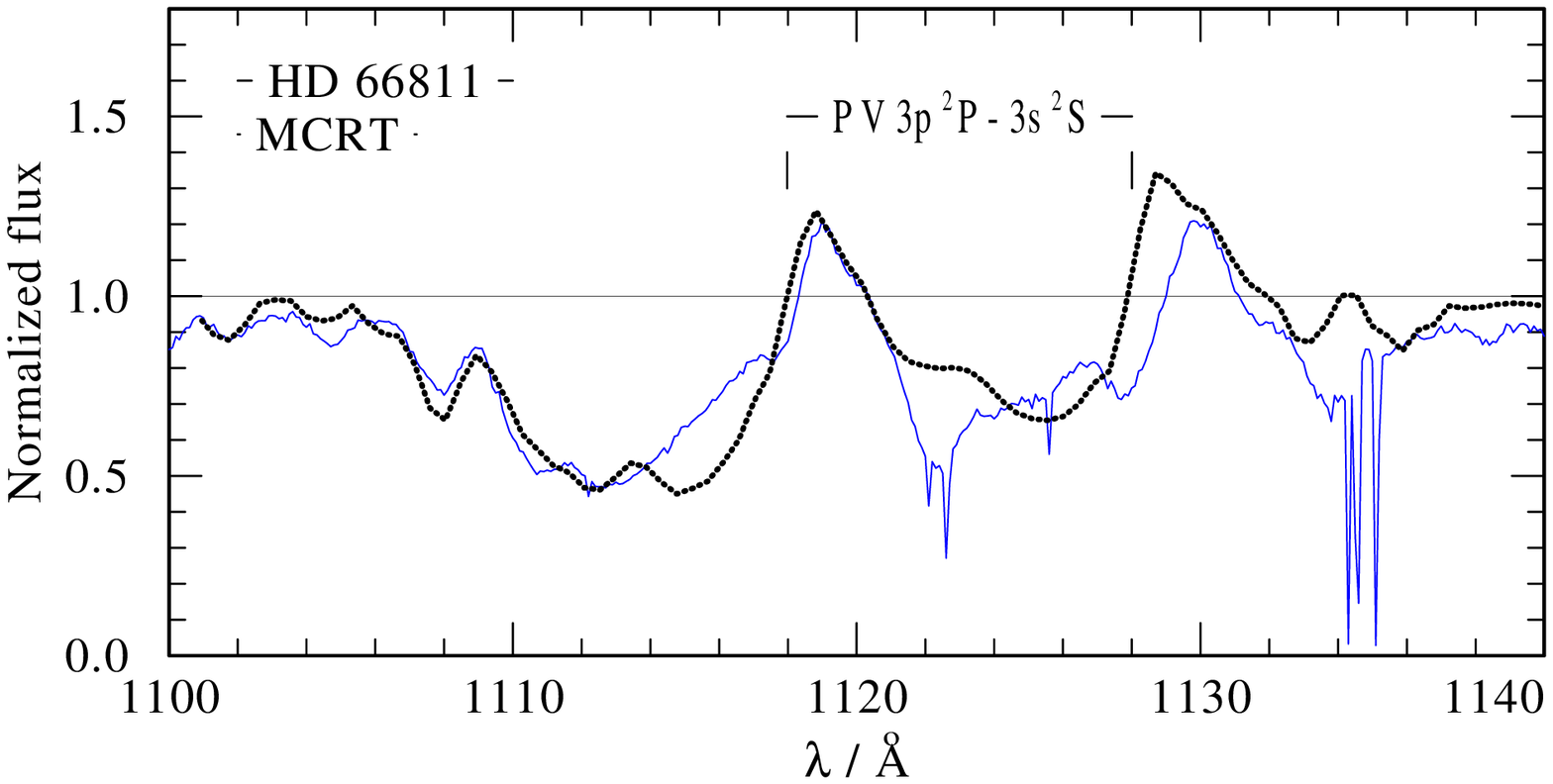}
\includegraphics[width=0.5\textwidth]{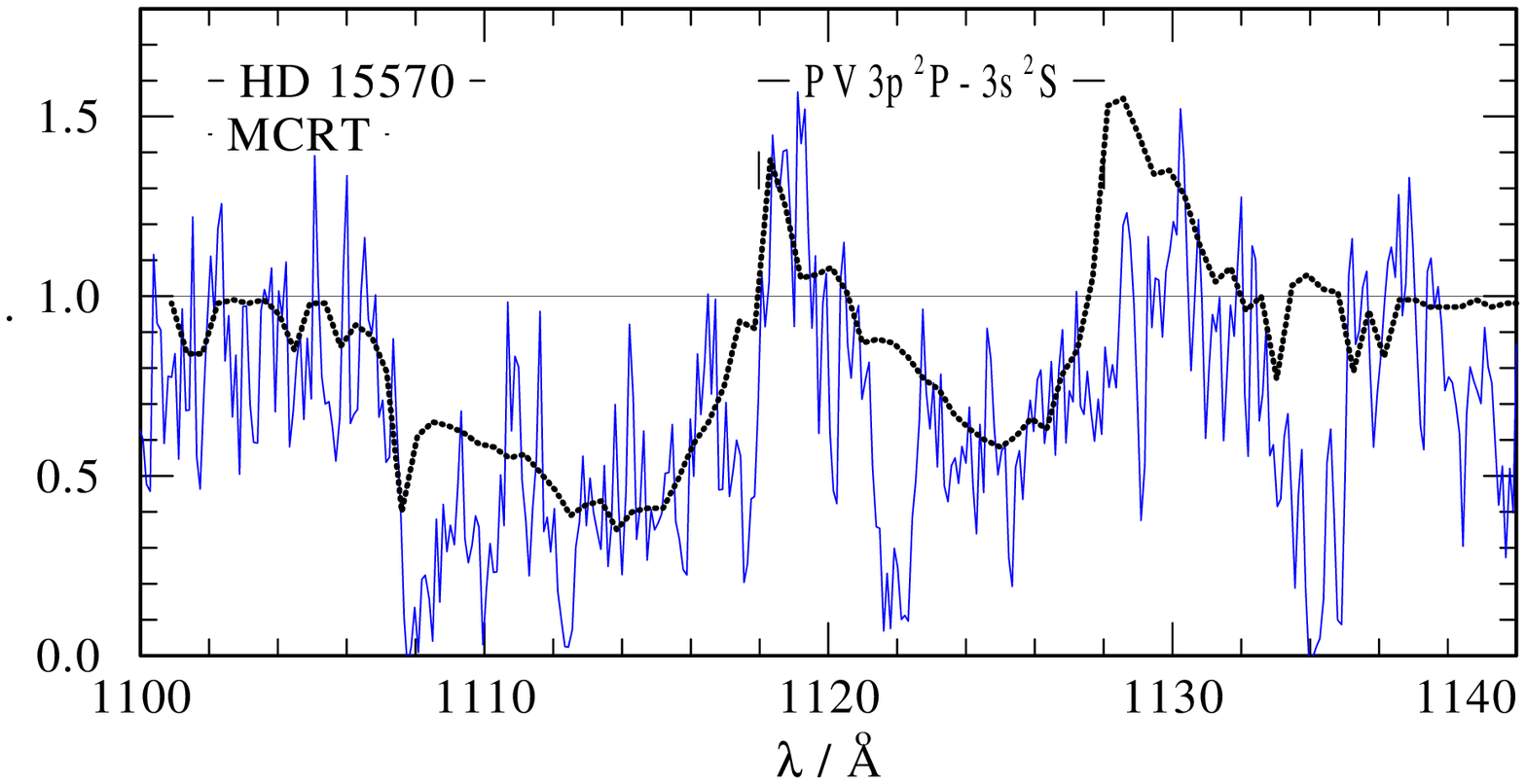}
\caption{The upper and the middle panels show the synthetic spectra
(dotted-black lines) of HD\,66811 (left panels) and HD\,15570 (right
panels) obtained from the {\POWR} models, i.e.\ without macroclumping.
Thin solid-blue lines are observed spectra. The dashed-green lines  in
the middle panels are from the same model, but only accounting for the
photospheric  lines while wind lines from {\ionp} and {\ionsi} are
suppressed. These photospheric spectra  are used as input for the
\mbox{3-D} Monte-Carlo calculations with macroclumping (lower panels).
The  parameters of these models are given in
Tables~\ref{tab:inputstelwindpara}-\ref{tab:clumpara}.}
\label{compone}
\end{figure*}

A set of parameters describes the inhomogeneous wind. The clumping
factor $D$ specifies  the density inside clumps in respect to the
smooth wind density. We used the same value as in the {\POWR} code for
the microclumping. Other clumping properties are the clump separation
parameter $L_{0}$, the density of the inter-clump medium $d$ (for the
case of a two component medium), 
and the radius {\rcl} where clumping sets on. The velocity range inside
each clump is described by the velocity deviation parameter
$m=\veldis(r)/\varv(r)$. For a more detailed description of these
parameters we refer to \cite{Surlan:EtAl:2012}.

\section{Model fitting}
\label{modcalc}

For each star from the sample, we perform the following 
procedure:

\renewcommand{\labelenumi}{\alph{enumi})}

\begin{enumerate}
\item
\mbox{1-D} models are calculated with the {\POWR} code in order to
establish the mass-loss rate from fitting the 
{\Halpha} line;
\item
then, the obtained mass-loss rate together with the {\ionp} ionization
fraction and the photospheric spectrum 
are  used as input for the \mbox{3-D} Monte Carlo simulations of 
the clumped wind. From optimizing the fit of the {\ionp} resonance
doublet, the clumping parameters are determined.
\end{enumerate}

These steps are described  in more detail in the next two subsections.

\begin{figure*}[t] 
\includegraphics[width=0.5\textwidth]{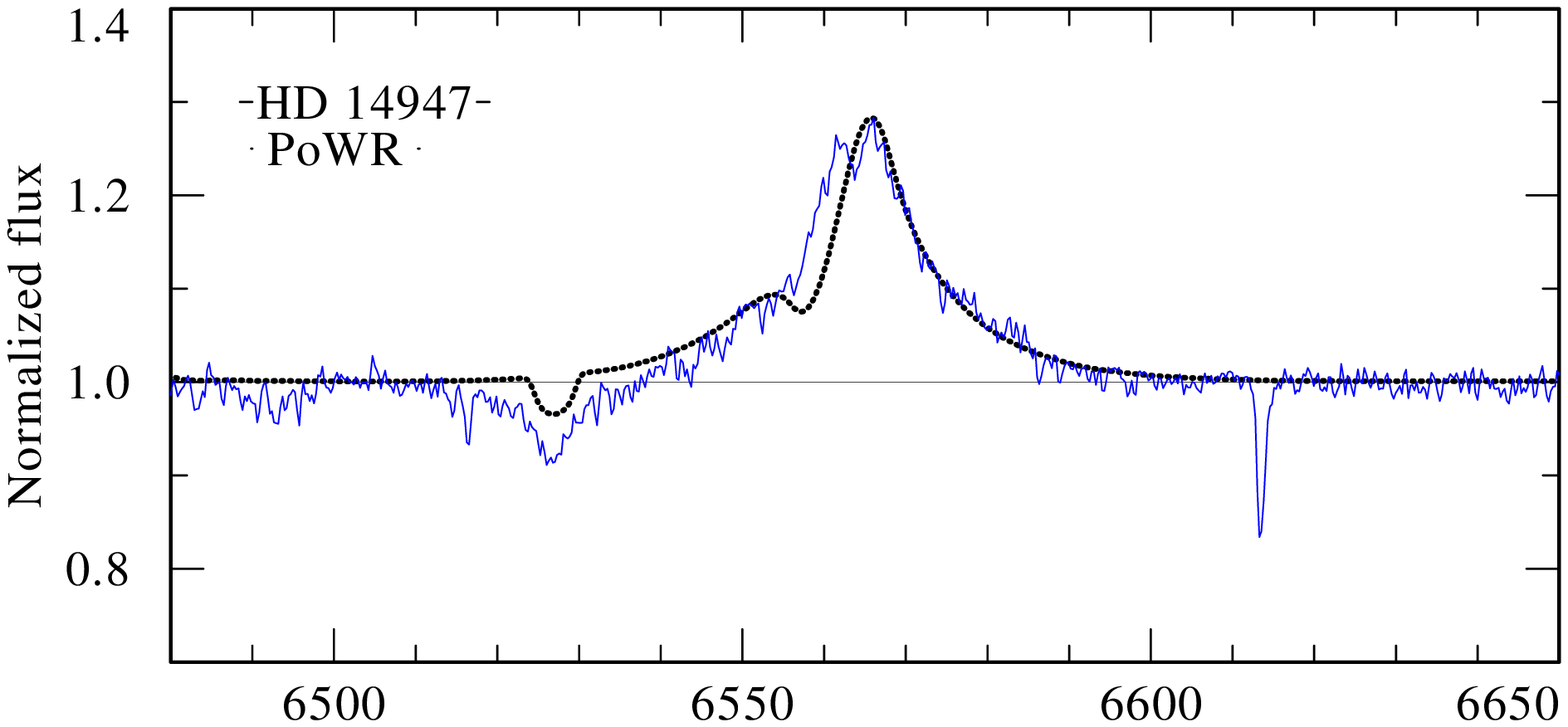}
\includegraphics[width=0.5\textwidth]{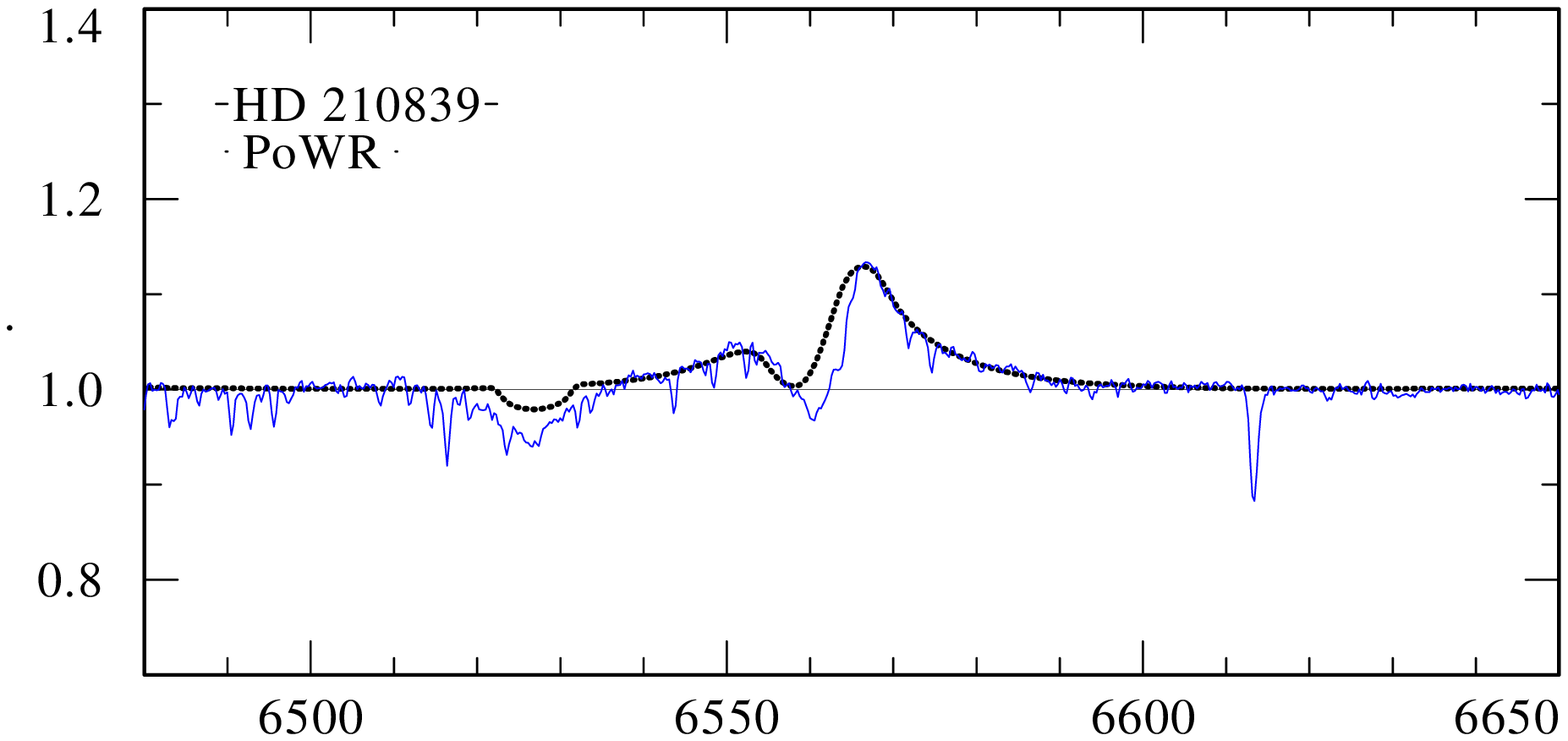}\\
\includegraphics[width=0.5\textwidth]{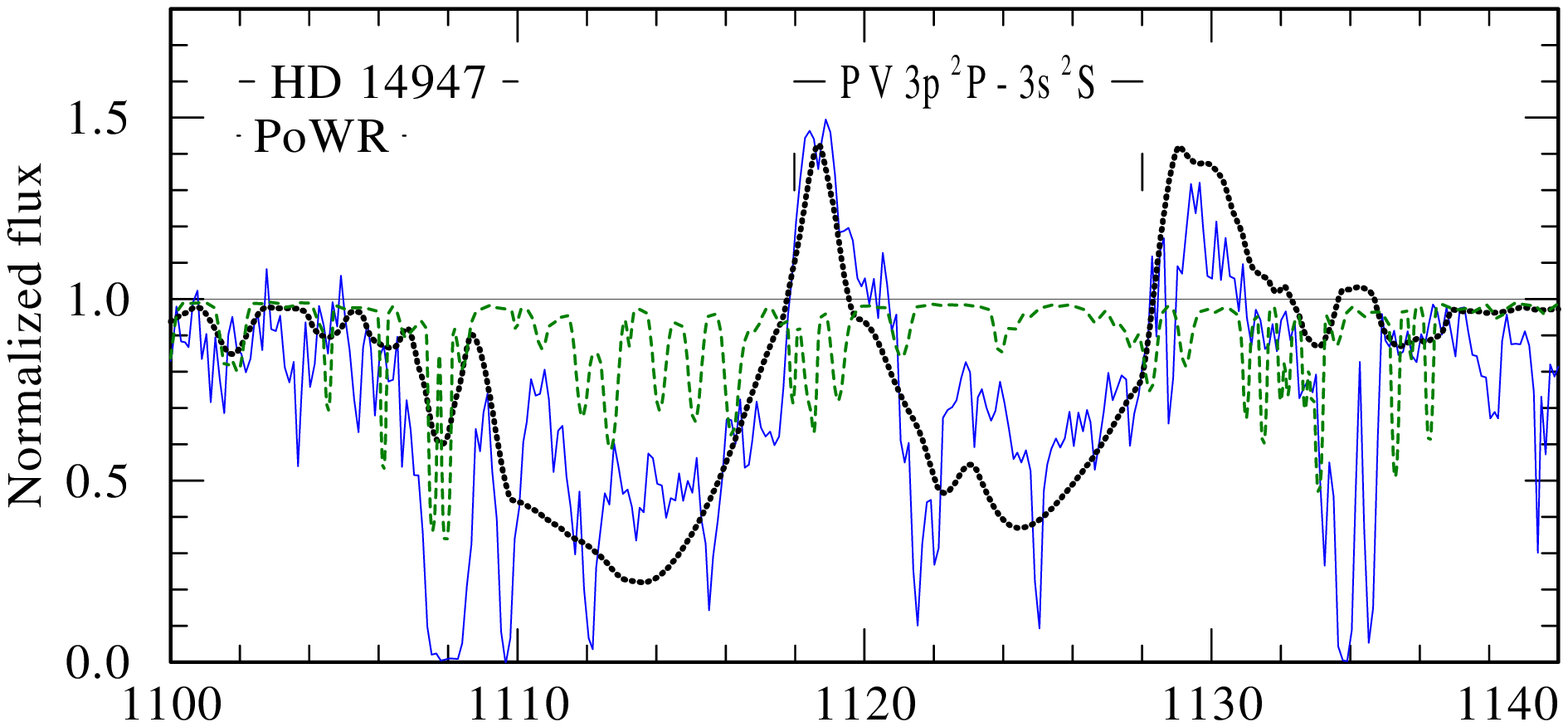}
\includegraphics[width=0.5\textwidth]{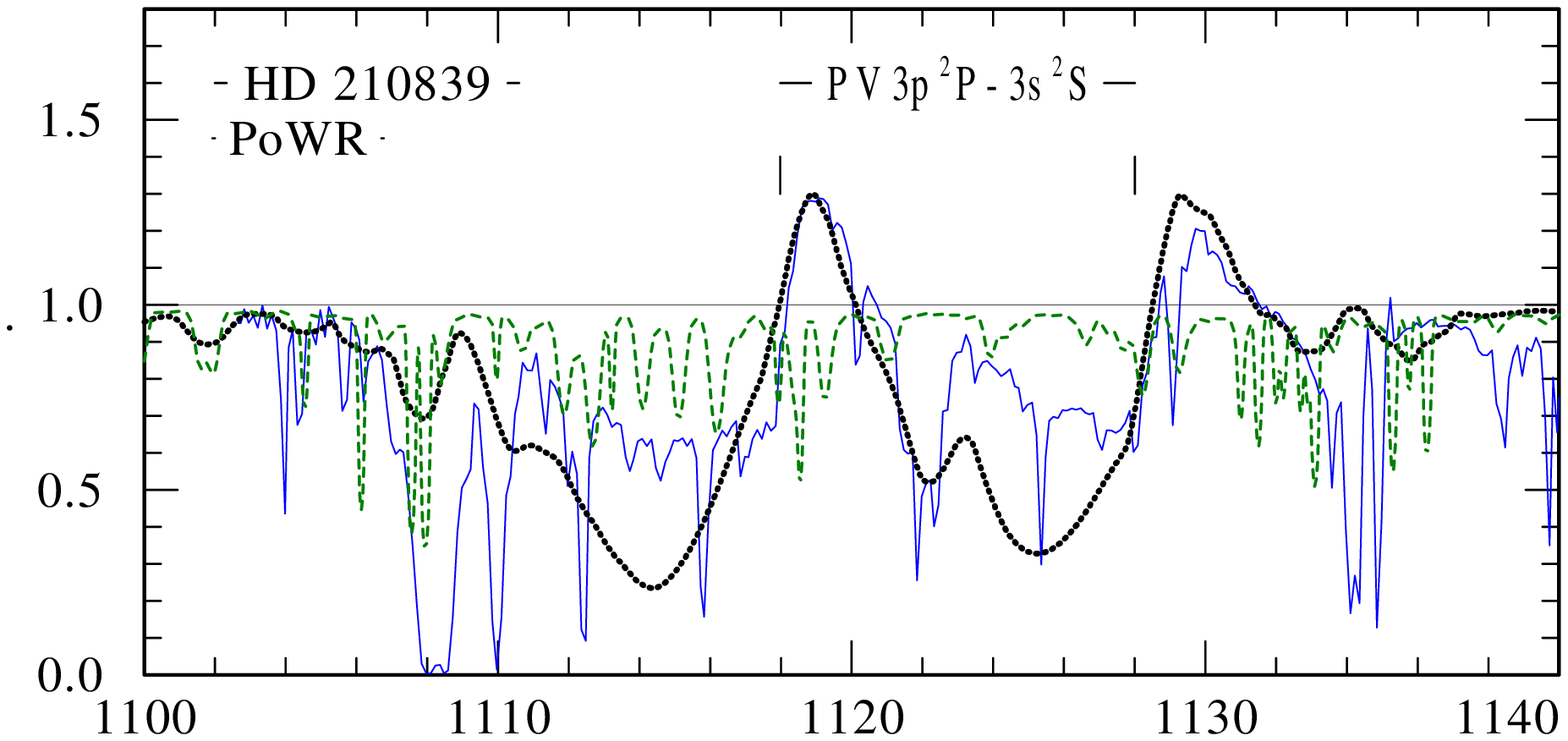}\\
\includegraphics[width=0.5\textwidth]{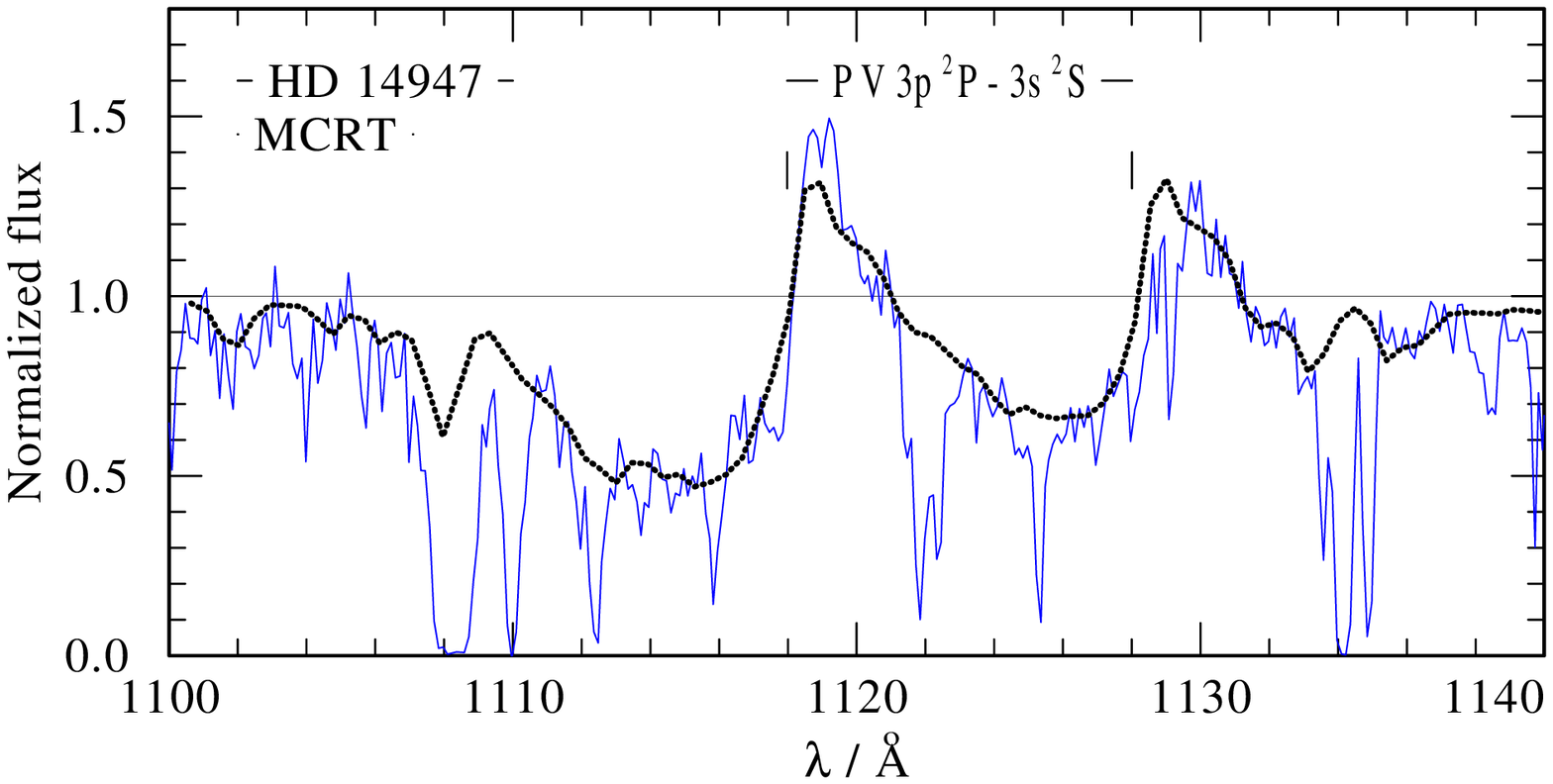}
\includegraphics[width=0.5\textwidth]{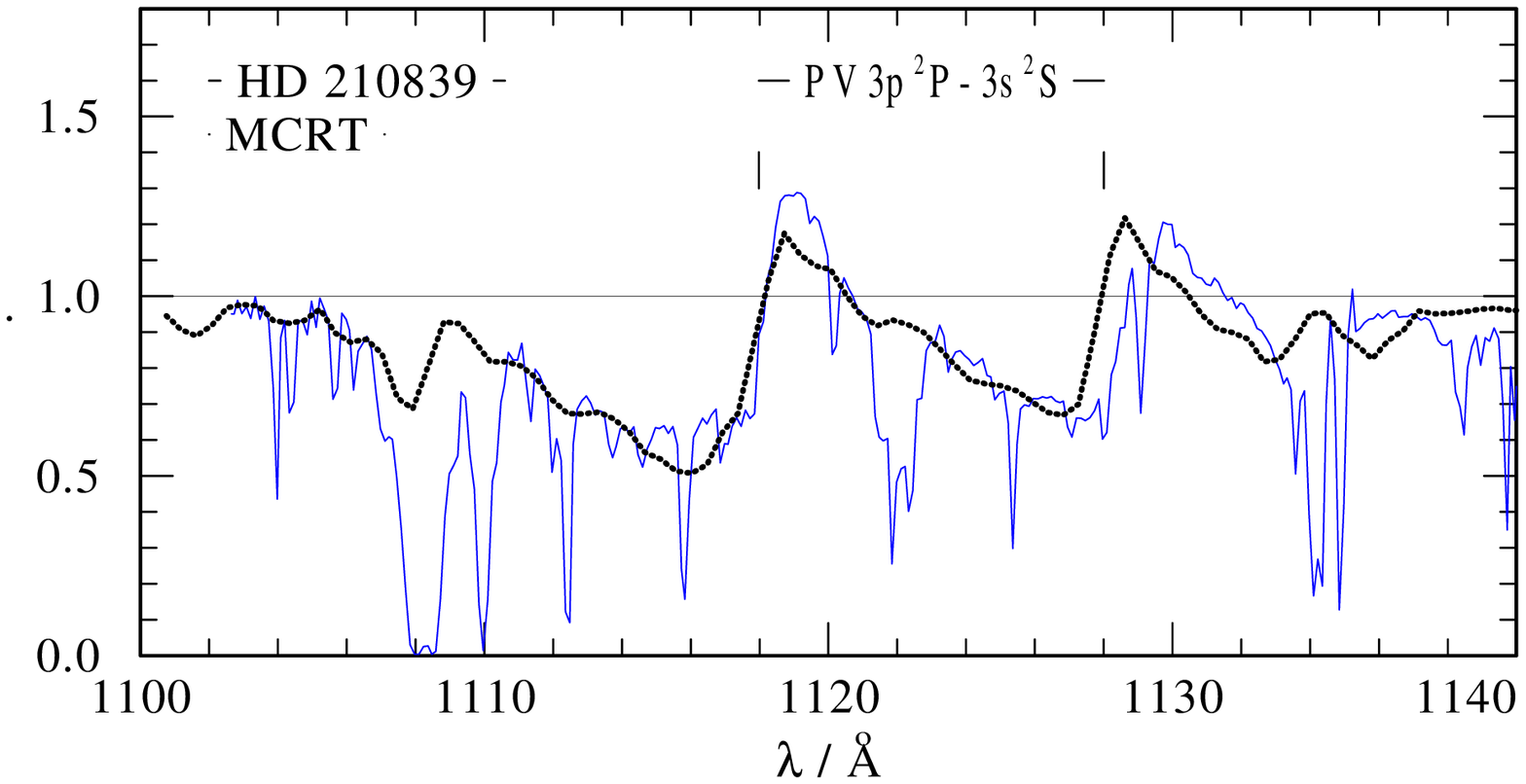}
\caption{Same as Fig.~\ref{compone} but for HD\,14947 (left panels) and
HD\,210839 (right panels).}
\label{comptwo}
\end{figure*}

\subsection{1-D {\POWR} model fitting}

As input to the \mbox{1-D} \POWR\ model calculations we used
the stellar and wind parameters and element abundances as compiled in
Table~\ref{tab:inputstelwindpara}. The mass-loss rates were slightly
adjusted in order to optimize the fit with the optical observations
({\Halpha}, {\Hbeta}, {\Hgamma}, and {\ionheii} lines). The finally
adopted \mdot\ are listed in Table~\ref{tab:derivedstelwindpara}.

All spectral fits are documented in the {\em Online Material}. Note that
the synthetic spectra were flux-convolved to simulate instrumental and 
rotational broadening, taking  $\varv\sin i$ from 
\cite{Bouret:EtAl:2012}. 

The UBVJHK photometry of all stars is taken from the GOS catalog
\citep{Maiz:EtAl:2004}. The color excess $E_{\rm B-V}$ is adopted from
\cite{Bouret:EtAl:2012}. We applied the reddening law from 
\cite{Cardelli:EtAl:1989} and adjusted the $R_{V}$ parameter to optimize
the fit between the Spectral Energy Distributions (SED) of the model 
and the flux-calibrated observations. Moreover, since we kept the
luminosity at the literature value, we   
adjusted the stellar distances to achieve the SED fits. 
Our final values for 
$R_{V}$ and the stellar distance are listed in Table~\ref{tab:derivedstelwindpara}. The
SED fits are documented in the upper panels of Figs.\,A.1-5 in the {\em
Online Material}.

\subsection{3-D Monte-Carlo model fitting}

\begin{table}[tb] 
\centering
\caption {Clumping parameters which give the best fit to the observed {\ionp} line profiles. 
All other model parameters are given in Table~\ref{tab:fixclumpara}.} 
\label{tab:clumpara} 
\small{\begin{tabular}{l c c} 	
\hline
\hline
\multicolumn{1}{c}{
\rule[0mm]{0mm}{3.5mm} \!\!\!
Star} &       $d$             &   $m$                     \\
\hline
\hline
\rule[0mm]{0mm}{3.5mm} \!\!\!
HD 66811                 &  0.15                 &   0.25           \\
HD 15570                 &  0.40                 &   0.20           \\
HD 14947                 &  0.20                 &   0.10           \\
HD 210839                &  0.15                 &   0.01           \\
HD 192639                &  0.10                 &   0.01           \\
\hline
\end{tabular}}
\end{table}

\begin{figure}[t]
\includegraphics[width=\columnwidth]{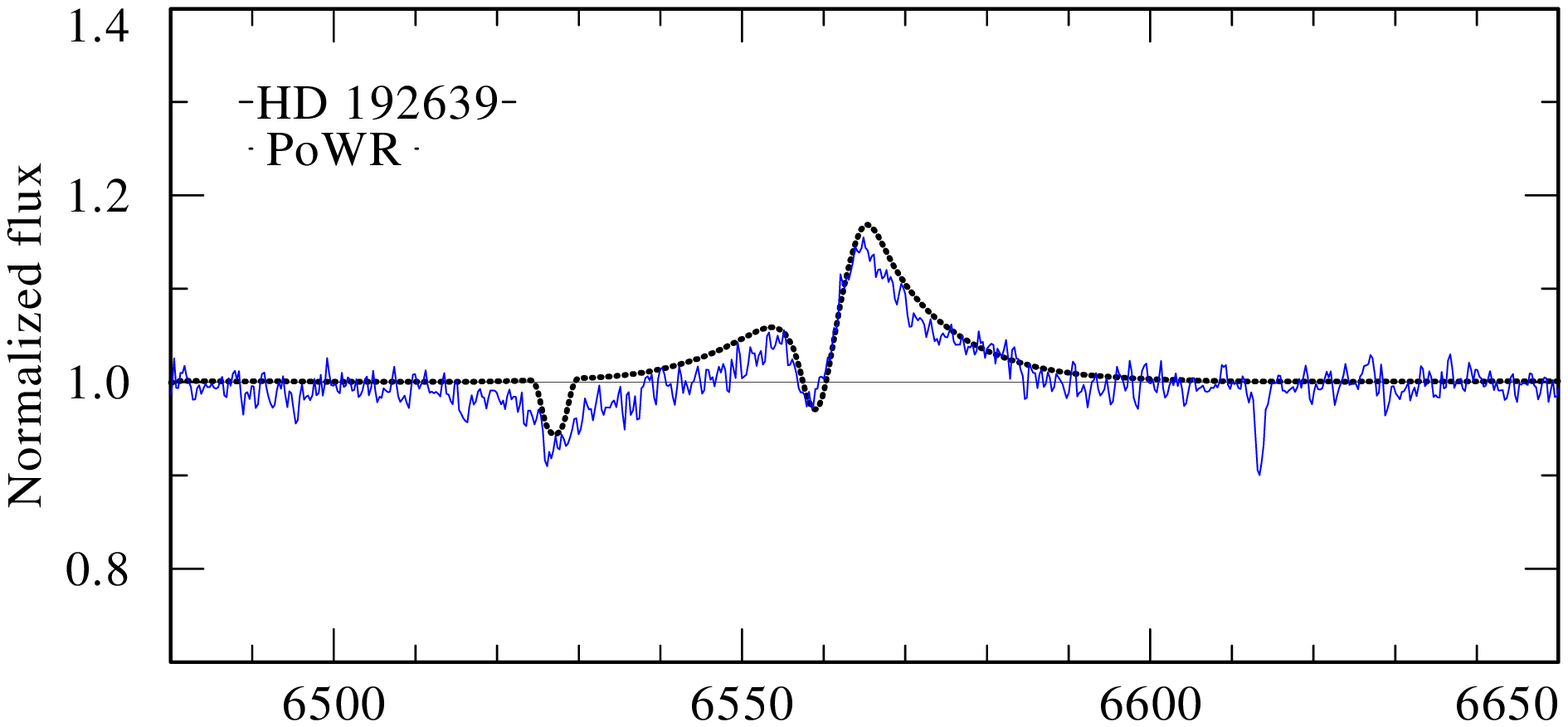}
\includegraphics[width=\columnwidth]{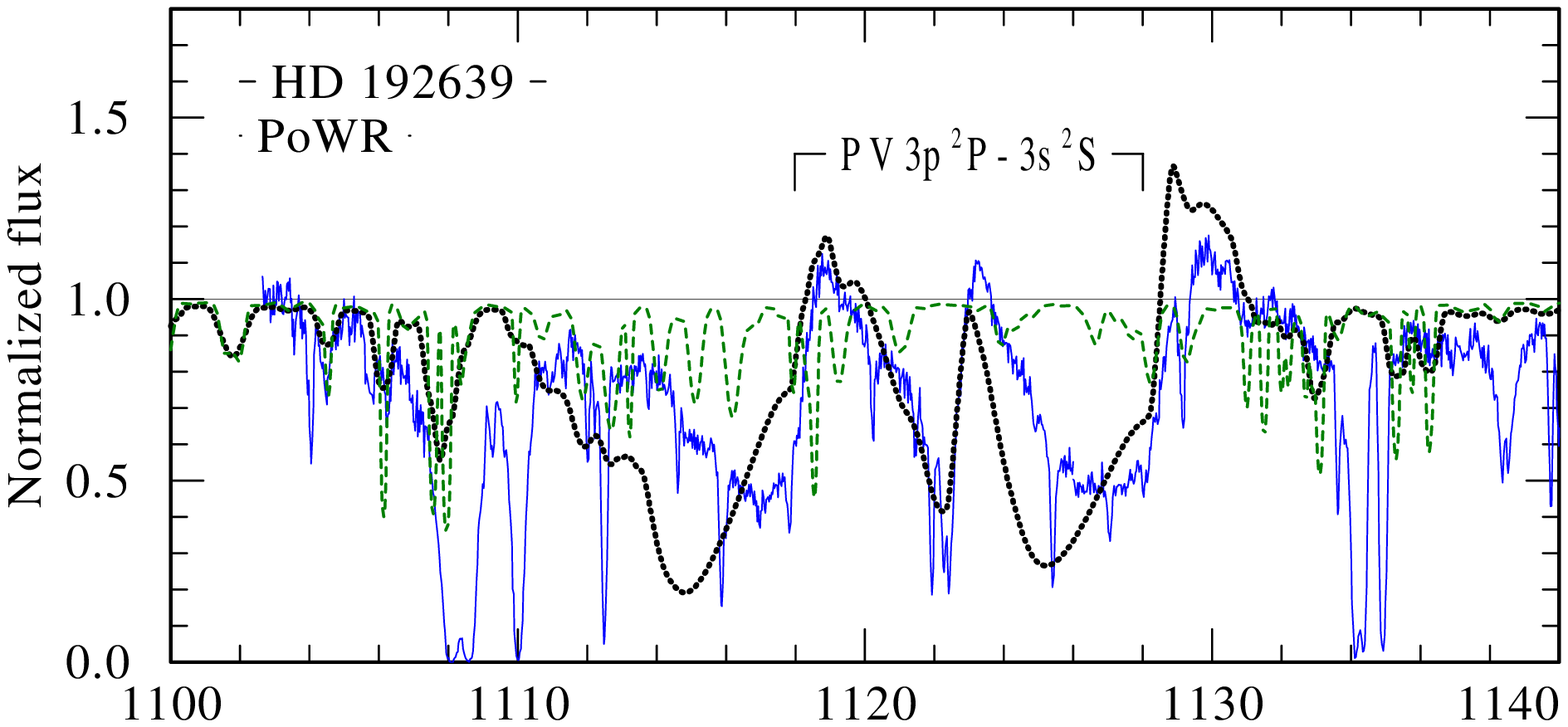}
\includegraphics[width=\columnwidth]{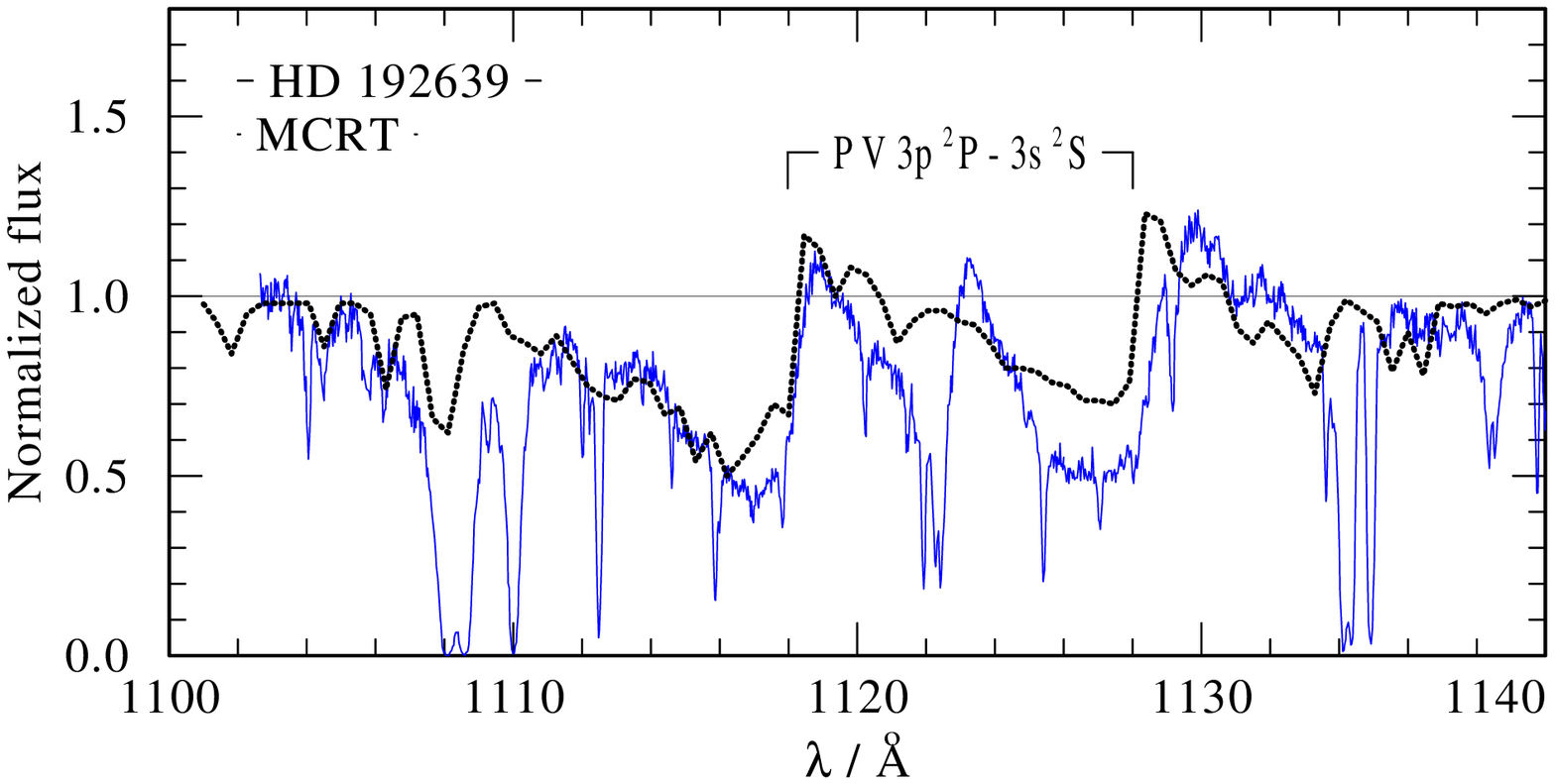}
\caption{Same as Fig.~\ref{compone} but for HD\,192639.}
\label{compthree}
\end{figure}

Once the mass-loss rate is established from the $H\alpha$
fitting, the stratification of the {\ionp} ionization fraction and
photospheric spectra is extracted from the final {\POWR} model, 
and used as input for calculating the \mbox{3-D} Monte-Carlo 
radiative transfer  in the clumped wind as described in
Section~\ref{tridmodel}. 

To be consistent to the {\POWR} models, the velocity of the wind 
is described by a {\dbetlaw} with the same parameters. The clumping 
factor $D$ and the Doppler-broadening velocity is taken 
consistent to the {\POWR} models.

While some of the clumping parameters are fixed (see
Table~\ref{tab:fixclumpara}), the inter-clump medium density factor
$d$ and the  velocity deviation parameter $m$ are varied in order to 
find the best fit to the observed {\ionp} doublet. Their final values  
are given in Table~\ref{tab:clumpara}.

\section{Results}
\label{results}

Let us first review the global results of the modeling. As it can be
seen from the upper panels of Figs.~\ref{compone}-\ref{compthree},  the
{\Halpha} line fits
reasonably well, although not perfectly
(see also {\Hbeta} and {\ionheii} profiles in {\em Online material}).
Comparing our fits with the fits obtained in other investigations, e.g.
by \cite{Bouret:EtAl:2012} for the star HD\,210839, our fits are not
worse.
Both our {\POWR} models and the models of \citeauthor{Bouret:EtAl:2012}
assumed microclumping throughout the wind.
The overall shape of the {\Halpha}, {\Hbeta}, and {\ionheii} 4686\,\AA\
lines is fitted well.


For an additional check of the fit, we compare the synthetic UV spectrum
with low-resolution IUE spectra (third panels in Figs.\ref{figzetpup} --
\ref{fighd192639}). All lines fit well except of the \ion{N}{v} resonance 
doublet. This line is partly formed in the tenuous inter-clump medium 
\citep{Zsargo:EtAl:2008} which is not included in the \POWR\ model 
calculations.

The {\ionp} resonance doublet, however, is
predicted far too strong by the \POWR\ models in all cases (see
Figs.~\ref{compone}-\ref{compthree}, middle panels). In contrast, excellent
agreement with observation is achieved with the \mbox{3-D}  Monte-Carlo
simulations (see Figs.~\ref{compone}-\ref{compthree}, lower panels).
The disagreement in the red component of the {\ionp} $\lambda\lambda~1118,
1128~${\AA} line for the star HD\,192639 (Fig.
\ref{compthree}) is caused by blending with the \ion{Si}{iv} 1128\,\AA\
line, which is not included in our \mbox{3-D} Monte-Carlo simulations.
%

In the following subsections we describe how we chose the clumping 
parameters for achieving the best agreement between calculated and 
observed {\ionp} profiles.
We demonstrate the effect of these parameters by
taking HD\,14947 as example.

\begin{figure*}[t]
\begin{centering}
\includegraphics[width=.8\textwidth]{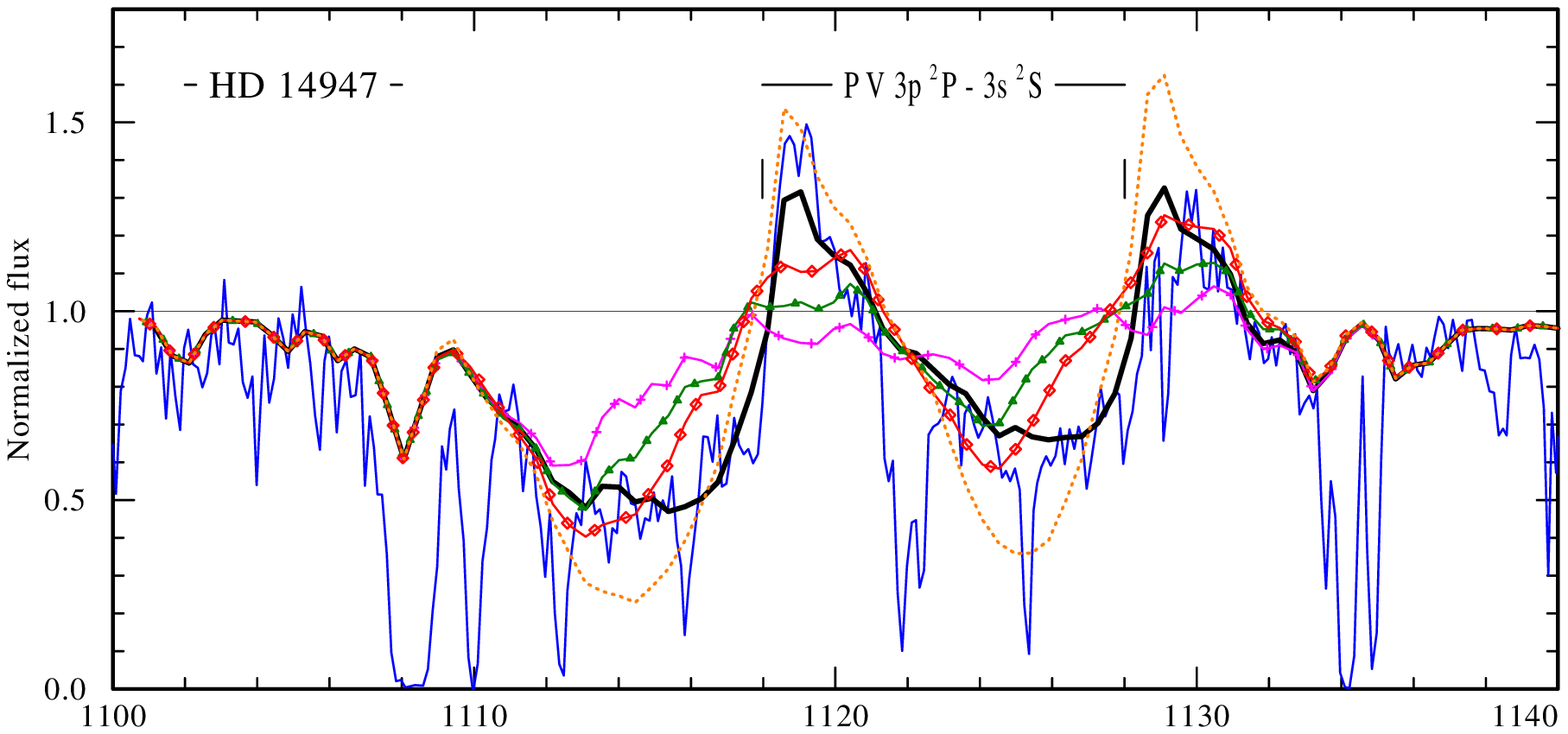}\\
\includegraphics[width=.8\textwidth]{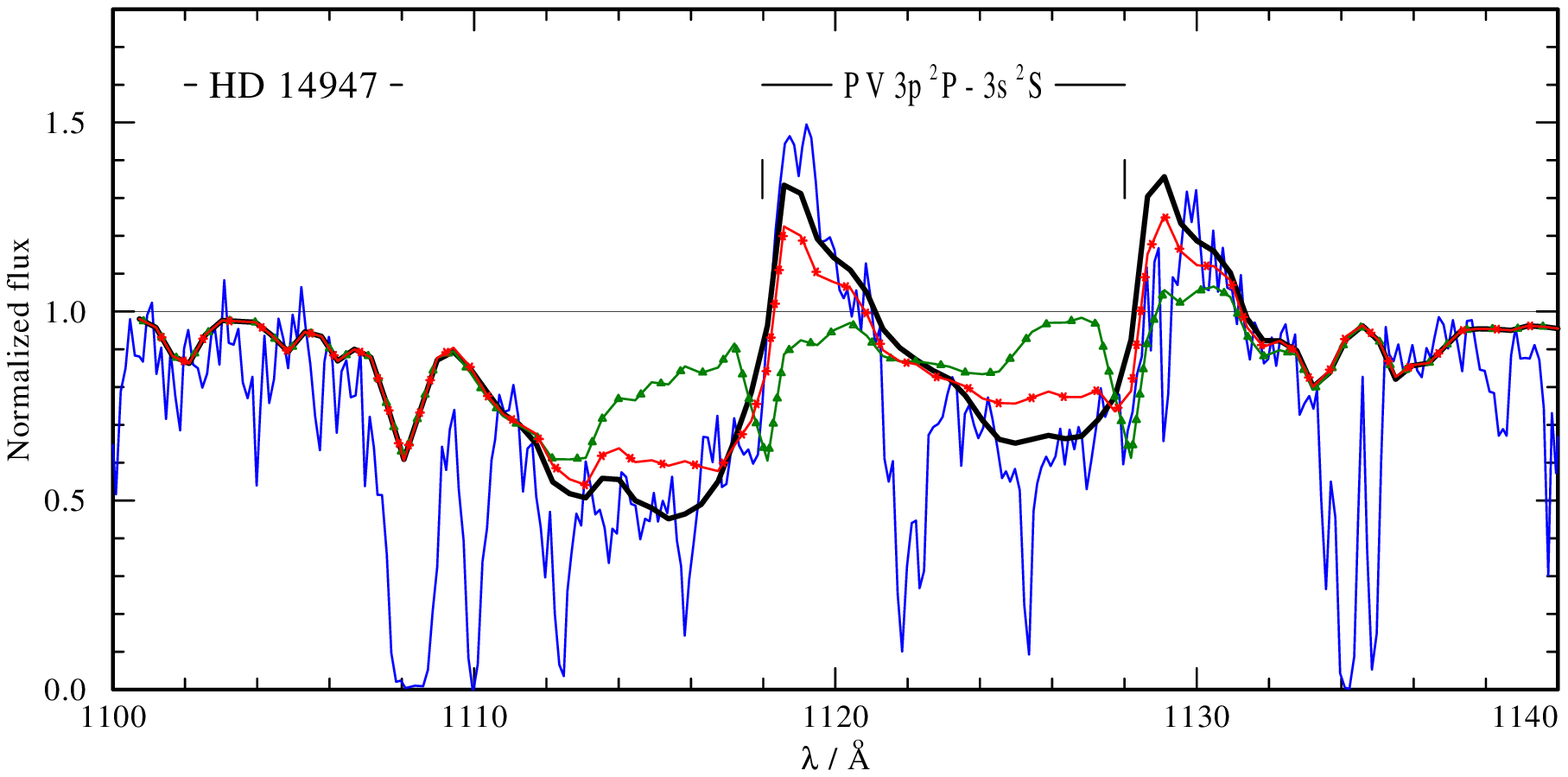}
\caption{Comparison between calculated {\ionp} and observed (thin
solid-blue lines) line profiles of HD\,14947. {\em Upper panel:} Effect
of the number of clumps. The purple line with crosses is calculated 
with $d=0$, $\rcl=1$, $m=0.1$, and $L_{0}=0.5$. The green line with
triangles and the red line with squares differ only by $L_{0}=0.2$ 
and $L_{0}=0.1$, respectively. The thick solid-black line is
calculated with $\rcl=1$, $m=0.1$, $L_{0}=0.5$, and $d=0.2$. The
dotted orange line is from the \POWR\ model and 
corresponds to a smooth wind. {\em Lower panel:}
Effect of the onset of clumping and the inter-clump medium density. The
green line with  triangles is calculated with $d=0$, $\rcl=1.1$,
$m=0.1$, and $L_{0}=0.5$. The red line  with asterisks differs only by
$d=0.1$, while the thick solid-black line is for $d=0.2$ 
and hence differs against the thick solid-black line in the upper 
panel only by the different $r_{cl}=1.1$. The remaining clump
parameters are fixed as given in Table~\ref{tab:fixclumpara}.}
\label{numbcl_icmeffects}
\end{centering}
\end{figure*}

\begin{figure*}
\begin{centering}
\includegraphics[width=.8\textwidth]{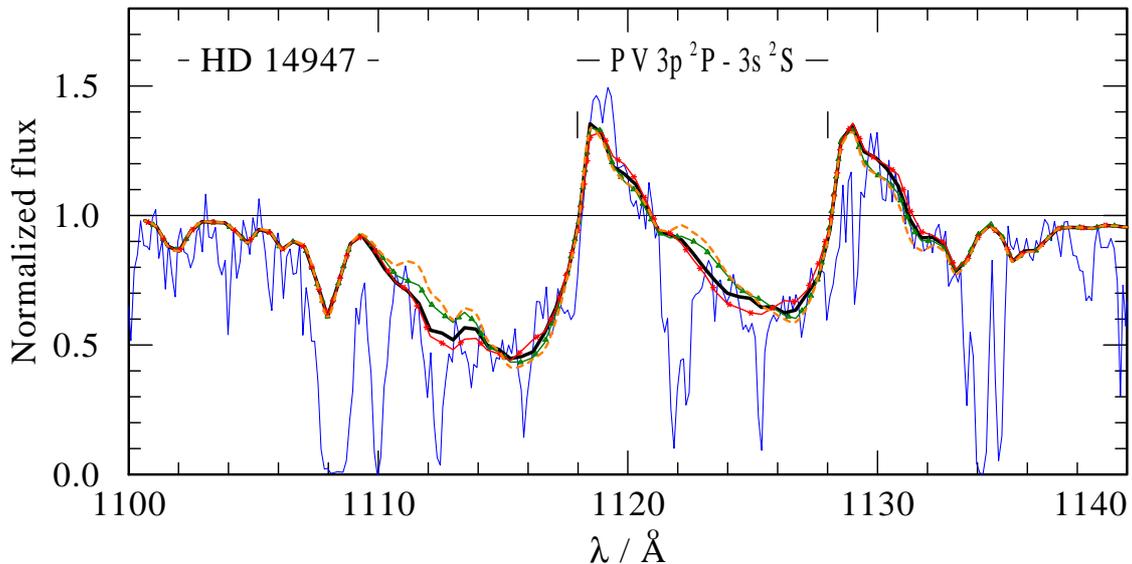}\\
\caption{Dependence of the {\ionp} profile on the clumping factor $D$. 
All four calculations are for $\rcl=1$, $L_{0}=0.5$, $d=0.2$, 
and $m=0.1$. The thick solid-black line is for our standard value $D=10$.
The red line with asterisks shows the profile for $D=5$,
the green line with triangles for $D=50$, and the dashed-orange line for $D=400$.
The thin solid-blue line is the observed spectrum.}
\label{Dvari}
\end{centering}
\end{figure*}

\begin{figure*}[t]
\begin{centering}
\includegraphics[width=.8\textwidth]{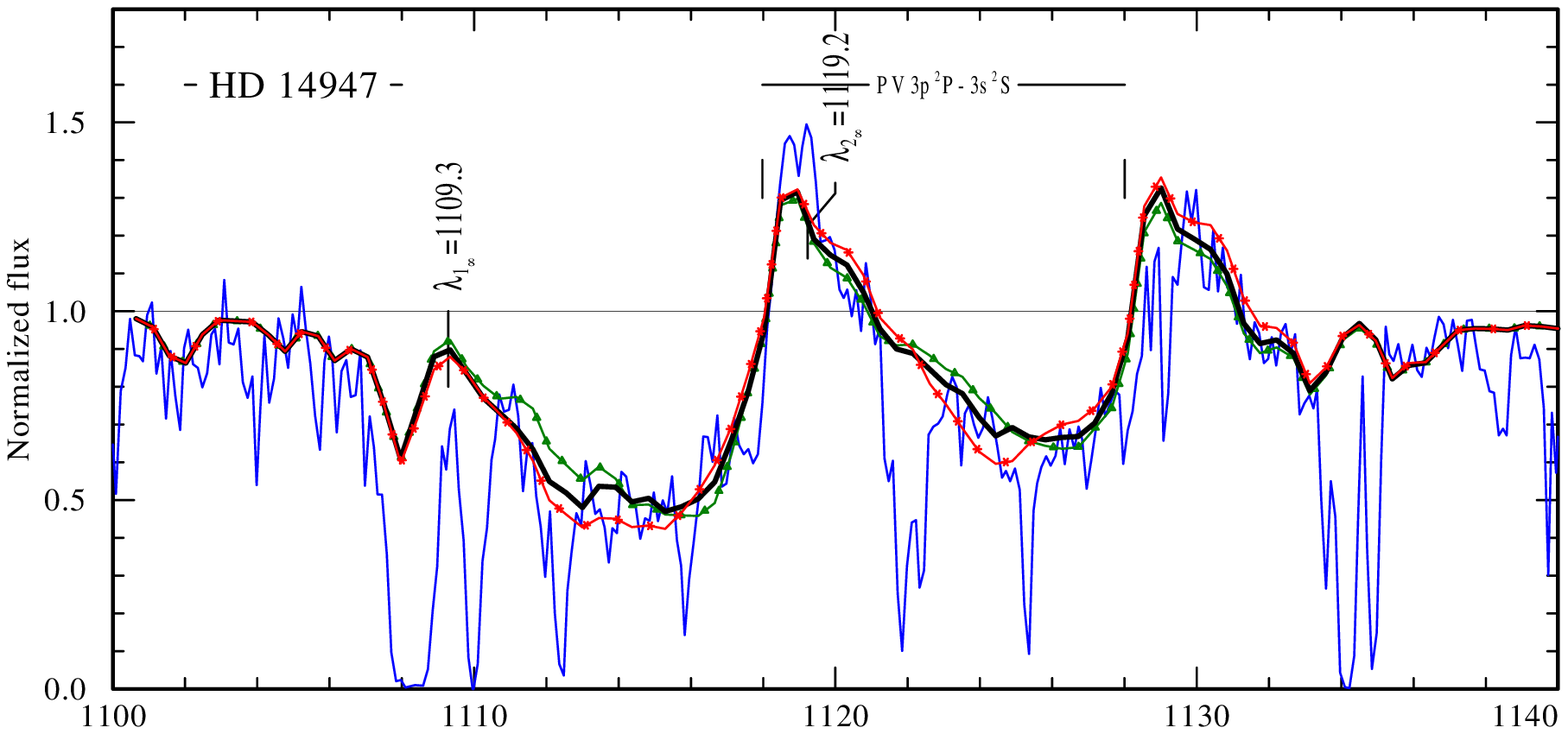}\\
\includegraphics[width=.8\textwidth]{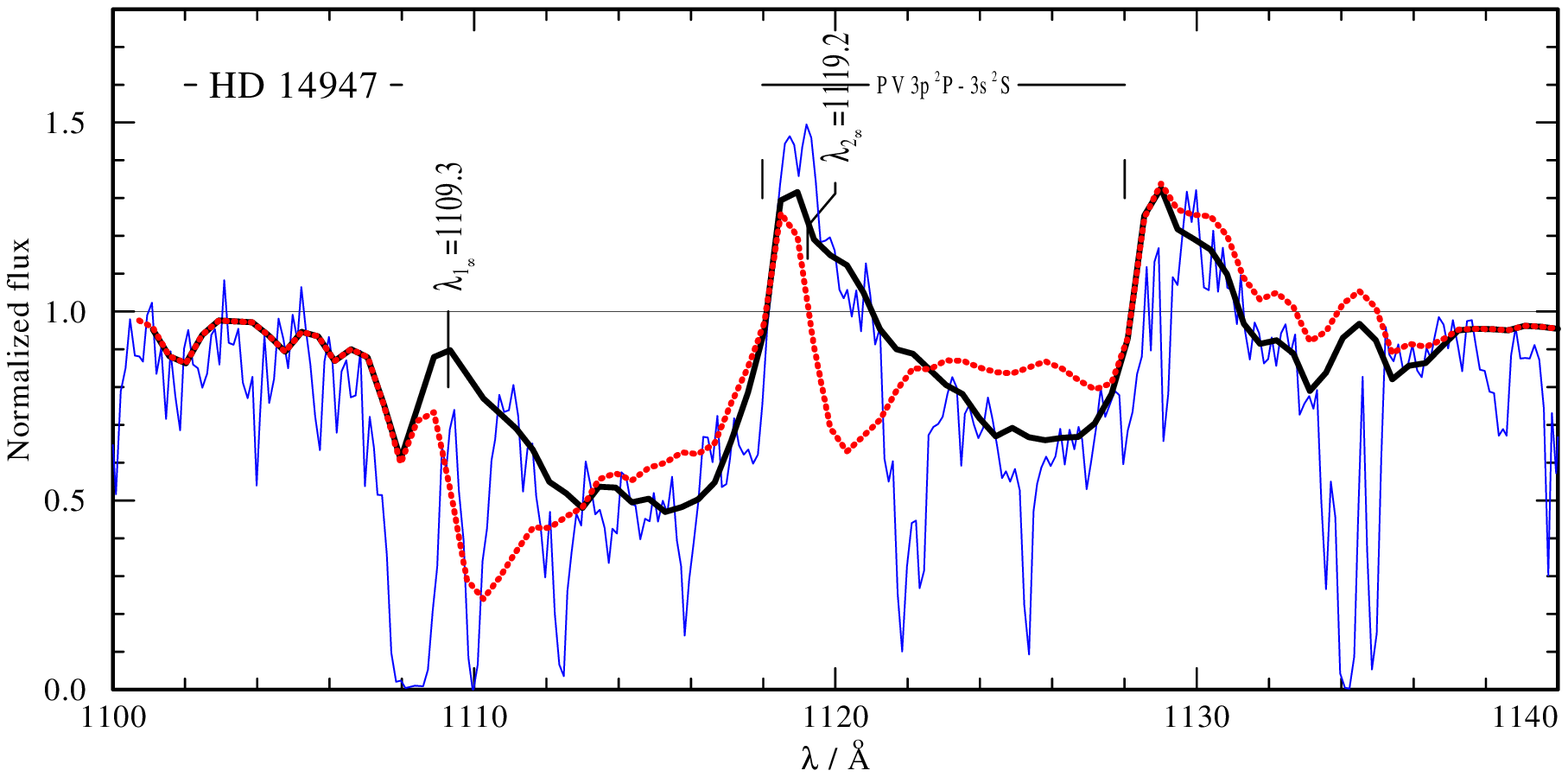}
\caption{{\em Upper panel:} The effects of the ``vorosity'' on the {\ionp}
profile. All three calculations are for $L_{0}=0.5$ and $d=0.2$. The green
line with triangles shows the profile for $m=0.01$, the red line with
asterisk for $m=0.3$, and the thick solid-black line for $m=0.1$. {\em
Lower panel:} {\ionp} line profiles calculated with standard
$\beta$-law and constant ionization fraction $q_{\ionp}=1$
(dotted-red line), compared to the simulation  
with {\dbetlaw} and the ionization stratification from the 
corresponding {\POWR} model (thick solid-black line).
Both profiles are calculated for $L_{0}=0.5$, $d=0.2$, $\rcl=1$, and
$m=0.1$. The thin solid-blue lines in the panels are the observed
spectrum. $\lambda_{1\infty}$ and $\lambda_{2\infty}$ represent the wavelength 
associated with the assumed \vinfty.}
\label{veldis}
\end{centering}
\end{figure*}

\subsection{Number of clumps}

The clump separation parameter $L_{0}$ controls the number of clumps,
{\ncl}, in the wind. Decreasing this parameter causes more clumps. For
very small $L_{0}$ ($L_{0}\rightarrow0$), the smooth wind is
approached (dotted orange line in the upper panel of
Fig.~\ref{numbcl_icmeffects}). It can be seen that with the smooth wind
approximation the absorption is deeper and the emission peaks are
higher. 
For a better fit with observation, we adopt the clumped-wind
model and set
$L_{0}=0.5$. This implies  $1.13\times10^4$ clumps within $100\,\R$ (cf.\
equation~24 in \citealt{Surlan:EtAl:2012}). 
The calculated line profile 
(purple line with crosses in the upper panel of
Fig.~\ref{numbcl_icmeffects}) is drastically reduced.

Now we increase the number of clumps, setting $L_{0}=0.2$ which implies
$1.75\times10^5$ clumps within $100\,\R$. Again, neither the strength of
the emissions nor the depth of absorptions can be reproduced (green line
with triangles in the upper panel of Fig.~\ref{numbcl_icmeffects}).
Even when we create as many as $1.4\times10^6$ clumps in the wind by
setting $L_{0}=0.1$ (the red line with squares in the upper panel 
of Fig.~\ref{numbcl_icmeffects}), the observed {\ionp} line  profile 
is not reproduced. 

One may compare these numbers with independent estimates  for the
number of clumps. E.g., \cite{Naze:EtAl:2013} recently found   that
more than $10^{5}$ clumps are required  in the wind of HD\,66811  in
order to explain its very low level of stochastic X-ray variability.

\subsection{Inter-clump medium density}
\label{icmdensity}

For a satisfactory fit  of the observed {\ionp} profile, additional
matter must be located between the clumps.  The inter-clump medium
density parameter $d$ \cite[see Sect.~2.1.2. in][] {Surlan:EtAl:2012}
defines its density. We find  that a reasonable fit to the observation
can be achieved this way, even with a lower number of clumps.  We set
$L_{0}=0.5$ and then increase $d$ until satisfactory agreement is
reached with about $d=0.2$ (solid-black line in the upper panel of
Fig.~\ref{numbcl_icmeffects}).

If we decrease $L_{0}$ (i.e.\ larger number of clumps), this
can be compensated by smaller values of $d$ to reproduce observation. Hence 
different combinations of $L_{0}$ and $d$ may give equally good
agreement with observation. From our clumped wind model it is not
possible to tell with certainty which combination of $L_{0}$ and $d$
corresponds to reality. We can only say that for winds which
consist of less than about $10^6$ clumps, inter-clump medium density is
a necessary ingredient of the wind in order to satisfactorily reproduce
the {\ionp} resonance doublet. But in any case, the inter-clump 
space cannot be void.

\subsection{Onset of clumping}
\label{onsetcl}

The parameter {\rcl} controls the radius where clumping sets on. Since
\cite{Sundqvist:Owocki:2013} showed that structures in the wind may
develop already very close to the wind base at $\rcl\lesssim 1.1\,\R$,
we check which effect the onset of clumping may have on the calculated line
profile.

First, we assume a one-component wind ($D=10$, $d=0$) and adopt the
$\rcl=1.1\,\R$. As a result, absorption dips appear close to the
laboratory wavelength of both {\ionp} doublet  components (green line
with triangles in the lower panel of Fig.~\ref{numbcl_icmeffects}). To
get rid of these sharp absorptions, we set the inter-clump density to $d=0.1$.
The result (red line with asterisks in the lower panel of
Fig.~\ref{numbcl_icmeffects}) shows that the absorption dips almost disappear,
but still the level of absorption is not fitted well. However,
after increasing the inter-clump density to $d=0.2$, the absorption
totally disappears, and the level of absorption is reproduced 
(solid-black line in the lower panel of Fig.~\ref{numbcl_icmeffects}).

The reason for this effect is that the inter-clump medium above {\rcl}
shields the lower, smooth part of the wind. If the former is dense enough 
(like for $d=0.2$), both absorption and emission are strong there to
hide the layers below. 

If we compare the solid-black lines in the upper and lower panels of
Fig.~\ref{numbcl_icmeffects}, which differ only by $\rcl$, we
cannot say which line fits the observations better, as both give
good agreement with observation. Even if we set
$\rcl=1.3\,\R$ and a bit different value of $d$, the agreement with 
the observed line remains similar. Hence, 
with appropriate values of $d$, the {\ionp} line profile can be
fitted equally well regardless if clumping starts from the surface of
the star or a bit above. The inter-clump medium hides the
spectral signature of the onset of clumping. Non-void inter-clump
medium has to be assumed always to fit the overall shape of the {\ionp}
profile.

For our final models presented in this paper we assume that clumping
starts at the base of the wind, for both the \POWR\ and the Monte-Carlo 
simulations.

%


%
\subsection{Dependence on the clumping factor $D$}

The parameter $D$ defines the density inside clumps with respect to the
smooth wind density. To check how different values of $D$ influence
formation of the {\ionp} resonance line profile, we varied this parameter
for one selected model (HD\,14947). The value $D=10$ gives a good fit to the
observation (see left panel of Fig. \ref{comptwo}). In Fig.~\ref{Dvari} we
now compare the profiles resulting for different values of $D$ between $5$
to $400$. While a slight preference exists for the fit with our standard
value $D=10$, the results depend only little on that parameter. In our
previous paper we demonstrated that enhancing the clump density parameter
$D$ leads to a more pronounced porosity effect \citep[cf.\ figure\,6
in][]{Surlan:EtAl:2012}. Since in this work we include the inter-clump
medium, this dependence is apparently reduced.

\subsection{Velocity dispersion inside clumps}

The porosity effect in the line radiation transfer depends on the
gaps in frequency between the line absorption of the individual clumps.    
This kind of porosity in the velocity coordinate is sometimes called 
``vorosity'' \citep{Owocki:2008}.
%
To investigate this
effect,
we study the impact of 
the velocity deviation parameter $m$ (see equation.~20 and figure\,3 in
\citealt{Surlan:EtAl:2012}), which allows the velocity inside the clumps
to deviate from the monotonic dependence on radius. As indicated by
hydrodynamic simulations, the velocity gradient 
is assumed to be negative there, while the ambient inter-clump medium is
moving monotonically according to the wind velocity law. 

The effects of $m$ on the {\ionp} line profile are shown in
Fig.~\ref{veldis}. From our modeling follows that ``vorosity'' mainly
affects the outer part of the wind, extending the absorption beyond
{\vinfty} and leading to a softening of the blue edge.  In all stars of
the sample, this improves the fit. However, we do not find any
significant reductions of the overall line strength. 
On the other hand, \cite{Sundqvist:EtAl:2010} found vorosity of similar
importance as porosity.
We can not confirm their conclusion.
%

\begin{figure*}[t]
\begin{centering}
\includegraphics[width=0.8\textwidth]{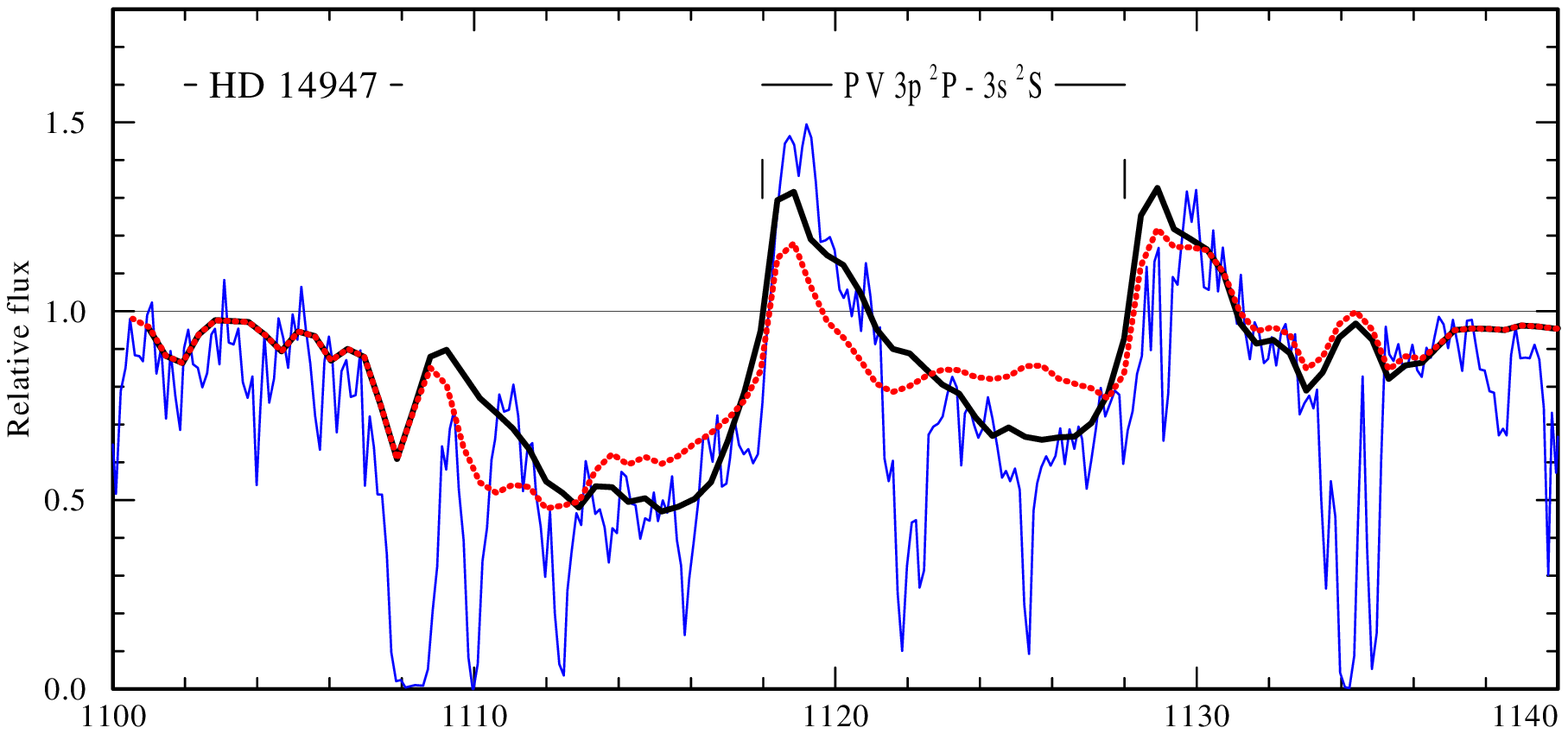}\\
\includegraphics[width=0.8\textwidth]{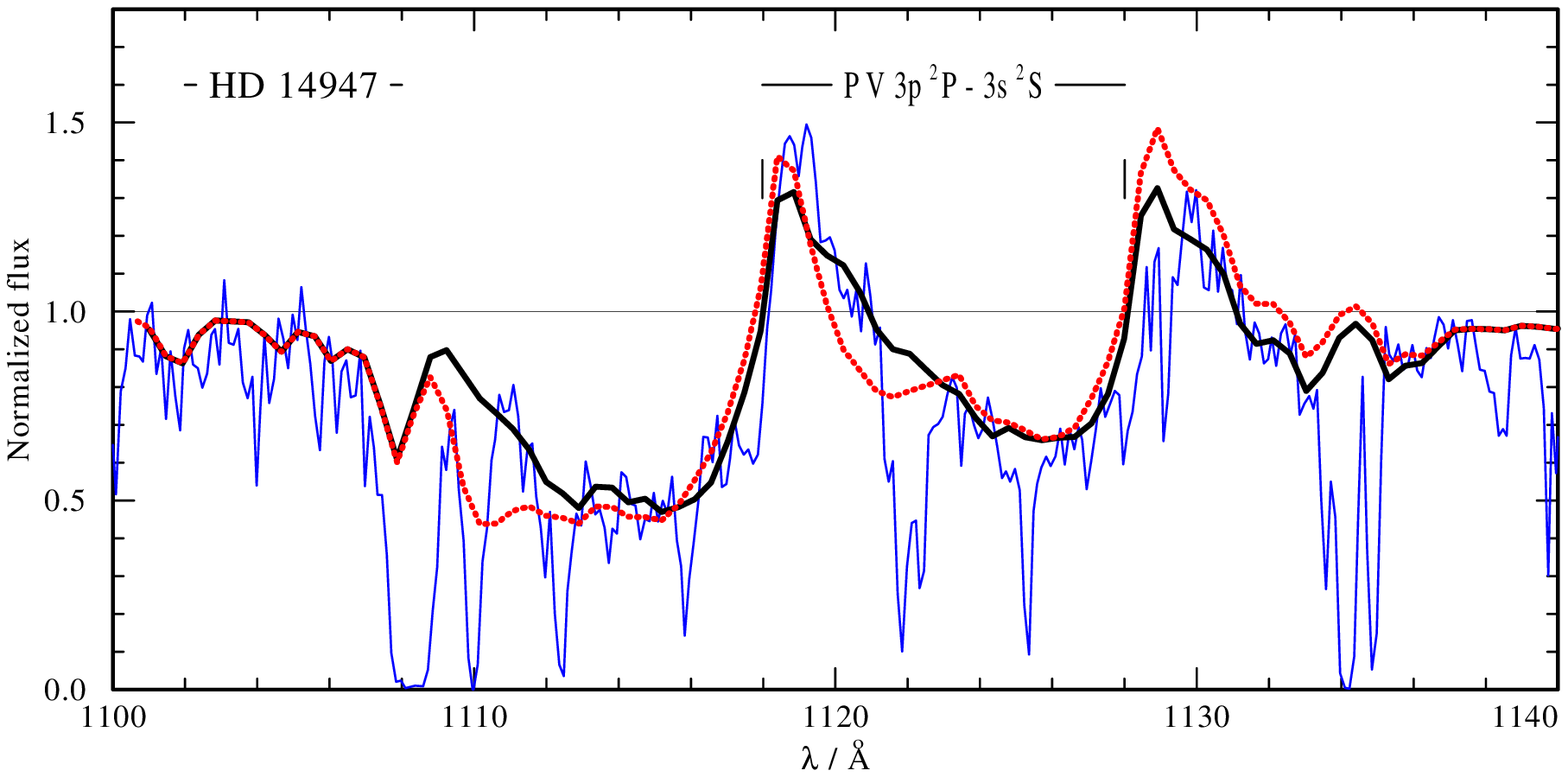}
\caption{Influence of velocity law and ionization fraction of {\ionp}
on profiles of the {\ionp} resonance doublet. Profiles are calculated
with $d=0.2$, $\rcl=1$, $m=0.1$,  and $L_{0}=0.5$ for the case of
HD\,14947. {\em Upper panel:} Comparison of profiles of the {\ionp}
doublet for standard $\beta$-law (the dotted-red line) and {\dbetlaw}
(solid-black line). {\em Lower panel:} Comparison of profiles of the
{\ionp} doublet for constant ionization fraction $q_{\ionp}=1$ (the dotted
red line) and ionization fraction following from the {\dbetlaw} (the thick
solid black line). The thin solid blue line is the observed spectrum.}
\label{beta-ionfrac-comp}
\end{centering}
\end{figure*}

\subsection{Clumping parameter degeneracies}
In our models, we are using several free parameters describing the
properties of clumping ($L_0$, $D$, $d$, \rcl, $m$).
However, it happens that for some different combinations of clumping
parameters we obtain similar or exactly the same {\ionp} line profiles.
There is a question, if it is possible to break such degeneracies of
parameters.
This is quite a complicated problem.
Using a single line it is probably impossible.
However, adding more lines to the analysis with different line opacities
and different depths of formation may put some additional constraints on
the adopted set of parameters.
For example, \cite{Zsargo:EtAl:2008} showed that the
\ionos~$\lambda\lambda~1032, 1038$ resonance doublet (created by Auger
ionization by X-rays from \ionoc) could be used to characterize the
inter-clump medium and break the degeneracy in the interpretation of
fits to the {\ionp} doublet.
However, their conclusion is based on 1-D wind model assuming
microclumping (consequently with less free parameters) and requires
verification for the more general case of a 3-D geometry.
It is an interesting problem, which goes beyond the scope of the present
paper, but we intend to study it in future by simultaneous fitting of
{\ionp} and {\ionos} doublets using their ionization structure obtained
from \POWR\ model, provided we find appropriate observational data.
Despite the fact that it is difficult to disentangle parameters of our 
model, our key conclusion is the accordance different mass-loss rate 
diagnostic when macroclumping is included in the models.
%

\subsection{Velocity law and {\ionp} ionization stratification}
\label{outerwind}

In our models we prefer the so called {\dbetlaw} (cf.\
Sect.\,\ref{sect:velfield}).  The main motivation for adopting the
{\dbetlaw} is to get rid of the  absorption dip close to the the blue
edge of the profile, which appears notoriously in models with the
standard $\beta$-law. With the {\dbetlaw}, the persistent acceleration
in the  outer wind enhances the velocity gradient there, and thus
reduces the line optical depth at largest blueshifts.  A physical
reason for such dynamic behavior  was suggested already by
\cite{Lucy:Abbott:1993}, who speculated that such a persistent
acceleration could arise from changes in the ionization structure. 
The mentioned blue edge absorption dip is clearly seen in 
Fig.~\ref{beta-ionfrac-comp} (upper panel),  where the dotted-red line
shows the calculation with the one-$\beta$-law, and the solid-black line
with the \dbetlaw.

In the lower panel of Fig.\,\ref{beta-ionfrac-comp} we demonstrate the
influence of the ionization stratification. The red-dotted profile is
calculated with   constant $q_{\ionp}=1$ in the whole wind, while for 
the thick solid-black line the ionization stratification has been taken
from the corresponding \POWR\ model.  For the parametric range of the
stars in our sample, the {\POWR}  models predict that more than $80\%$
of phosphorus is in ionization stage of {\ionp}, except close to the
photosphere. The {\dbetlaw} was employed in both cases. 
As can be seen,  the decrease of the {\ionp} ionization fraction
results in a shallower absorption in the outer wind.

\section{Discussion}
\label{diskuse}

\subsection{Clumping in the inner wind}

Time-dependent 1-D hydrodynamic models of radiation-driven winds 
always predicted that the line de-shadowing instability grows only in
the  acceleration zone. Thus, strongly inhomogeneous structures were
expected  to develop only at radii $\rcl \gtrsim 1.3\,\R$
\citep{Feldmeier:EtAl:1997, Runacres:Owocki:2002, DessartOwocki:2005}.
Recently, however, \cite{Sundqvist:Owocki:2013}  obtained structures
already close to the wind base ($\rcl \lesssim 1.1\,\R$), when they
accounted for the effect of limb darkening. 


There are observational arguments to suggest that
%
clumping may start close to the stellar surface
\citep[e.g.][]{Puls:EtAl:2006}. 
The early onset of clumping close to the stellar photosphere may 
be a consequence of the subsurface convection as investigated by 
\cite{Cantiello:Braithwaite:2011}. The presence of X-ray emission
very close to the photosphere \citep{Waldron:EtAl:2007} is an 
argument to support these considerations.

In our last paper \citep[][]{Surlan:EtAl:2012} we also concluded  that
clumping sets on close to the wind base, because otherwise deep,
un-shifted absorption features should be seen in the \ionp\ resonance
doublet which are not observed. In the light of our present results,
this conclusion is not so firm anymore. As mentioned in
Sect.~\ref{onsetcl}, the observations can also be reproduced if we assume
$\rcl=1.3\,\R$, while setting the inter-clump medium parameter to $d=0.2$. 
Obviously, the inter-clump medium reprocesses the radiation coming from
the base of the wind. Hence it is not possible to tell if the lower
wind is already clumped or not, at least not for the relatively dense
supergiant winds investigated here.

\subsection{Clumping in the outer wind}

The hydrodynamic simulations of the line de-shadowing instability
mentioned above predict that the clumps persist to large distances 
from the star. This justifies
our assumption that clumping extends to large radii, even to $\rmax
\gtrsim 100\,\R$.

Smooth-wind models predict much deeper absorption in the blue part of
the line profile than observed  (see Fig.\,\ref{veldis}). Obviously,
the effective opacity in the outer part of the wind is over-estimated.
Most importantly, we showed that this opacity can be effectively
reduced by the porosity effect. Additionally, the correct ionization
fraction of {\ionp} lowers the absorption a bit, and the {\dbetlaw}
distributes the line opacity  more uniformly across the line profile. 

Our \mbox{3-D} Monte-Carlo radiative transfer calculations were
performed with constant clumping parameters throughout the wind, which
implies that the clumps become larger and more separated from each
other with growing distance from the star. We should note here that
the clumping parameter $D$ may in fact be depth dependent
\citep[see][]{Puls:EtAl:2006}.

\subsection{{\ionp} ionization fraction}

The importance of the {\ionp} ionization fraction was pointed out by
\cite{Crowther:EtAl:2002}. The studies which had stated the mass-loss
rate discrepancy \citep{Massa:EtAl:2003, Fullerton:EtAl:2006} have used
simplified methods for simulating the P-Cygni profiles of the \ionp\ 
resonance doublet, namely the ``Sobolev with exact integration'' (SEI)
method  \citep{Lamers:EtAl:1987}. For this approach, the ionization
fraction of the \ionp\ ion must be adopted, and has been assumed to be
constant throughout the wind and close to unity (i.e., all phosphorus
is in the \ionp\ ground state). Our detailed non-LTE models show that
the  ionization fraction of \ionp\ is actually somewhat lower, and can
vary with radius.

Additionally, the ionization fraction of {\ionp} may be affected by
X-rays. This influence was examined by \cite{Krticka:Kubat:2009} who
showed that X-rays of the observed intensity cannot deplete the {\ionp}
ionization fraction significantly. Still, \cite{Waldron:Cassinelli:2010}
suggested that strong emission line radiation in the XUV energy band
can significantly reduce the abundance of {\ionp} and thus explain the
discrepant low mass-loss rates estimates. However, 
\cite{Krticka:Kubat:2012} showed that if the XUV radiation would be
strong enough to reduce the ionization fraction of {\ionp}, it would
also change the ionization balance of other elements and significantly
reduce the wind driving force, consequently also stellar mass-loss
rates.

Here we performed tests by including an X-ray field into the \POWR\
calculations, using the parameters of X-ray radiation as obtained 
from observations \citep{Oskinova:EtAl:2006}.
. We confirm the result found by \cite{Krticka:Kubat:2009}
that the X-rays really have no effect on the ionization balance of
phosphorus, especially on the abundance of {\ionp}. Similar conclusions
were also drawn by \cite{Bouret:EtAl:2012}.

\subsection{Mass-loss rates}

In principle, mass-loss rates through radiatively driven stellar winds
can be predicted from adequate hydrodynamical models.  Ideally, such
models would yield \mdot\ and \vinfty\ from a given set of stellar
parameters (stellar luminosity, mass, radius, and chemical
composition). Practically, such codes have to accept severe
approximations. In most of them, the radiative force is 
parameterized using the so-called force multipliers 
\citep[see][]{Castor:EtAl:1975,Abbott:1982}. Some of these codes
calculate the radiative force in detail from a list of  spectral lines
\citep{Krticka:Kubat:2004,Krticka:Kubat:2009,Krticka:Kubat:2010}.

Hydrodynamical stellar wind models which account for 
clumping are still missing. In a few test calculations, 
\cite{Krticka:EtAl:2008} and \cite{Muijres:EtAl:2011}
studied the effect of clumping on the radiative force.

\cite{Vink:EtAl:2000} performed Monte-Carlo calculations of the
radiative force, also using detailed line lists. However, they 
assumed the velocity law, instead of a fully consistent 
hydrodynamical solution. Conveniently, they condensed their results into 
a fit formula, which is widely used as reference for mass-loss rates.

Therefore we also compare our mass-loss rates, which are consistent
with the H$\alpha$ emission {\em and} the un-saturated UV resonance
doublet of \ionp, with the Vink formula
(Table~\ref{tab:derivedstelwindpara}, last column). On the average, our
mass-loss rates are smaller by a factor of two.

However, one should keep in mind that our \mdot\ relies on the
assumption that the clumping contrast is $D=10$. Within some range of
$D$ a simultaneous fit of H$\alpha$ and \ionp\ may be possible as well,
with somewhat different mass-loss rates  $\dot{M} \propto D^{-1/2}$.

%
%
%
%



\section{Summary}
\label{conclusion}
 
For a sample of five O-type supergiants, we studied the effects of
wind clumping on the mass-loss rate determination, considering
simultaneously the {\Halpha} emission (and other Balmer and {\ionheii} 
lines) and  the un-saturated resonance doublet of {\ionp} in the
far-ultraviolet. 

\begin{itemize}

\item 
When accounting for macroclumping,  it is
possible to simultaneously fit the {\Halpha} and the {\ionp} lines
with the same mass-loss rates. 

\item 
The consistent fit is achieved when we simulate the \ionp\ resonance
profile with our 3-D Monte-Carlo code for the line radiation transfer 
in clumpy stellar winds.  Obviously, the reported discrepancies between
{\Halpha} and {\ionp} mass-loss rates were due to the inadequate
treatment of clumping. 

\item 
The mass-loss rates for our consistent fits are lower by a factor of
1.3 to 2.6, compared to the mass-loss formula by \cite{Vink:EtAl:2000}. 

\item 
In contrast to other studies, it was neither necessary to reduce the
mass-loss rate by adopting an extremely high degree of clumping, nor
to assume a sub-solar phosphorus abundance for our consistent fits.  

\item 
The porosity that is needed to fit the \ionp\ lines implies that 
$\sim10^4$ clumps populate the wind within $100\,R_\ast$ at a given
moment. 

\item 
The velocity dispersion inside the clumps has a moderate effect on the 
porosity of the wind, and hence on the \ionp\ profile. The smaller this
dispersion, the smaller is the effective line opacity.  

\item
Compared to the standard $\beta$-velocity law, the double-$\beta$ law 
improves the detailed fit of the \ionp\ line profile. 
It smooths the blue absorption edge and removes the absorption 
dip close to that edge. 

\item 
With the detailed ionization stratification of \ionp\ from the {\POWR}
code, a better agreement with observed {\ionp} line profile can be achieved
than with   $q_{\ionp} \equiv 1$.

\end{itemize}

Our results emphasize that an adequate treatment of the line formation 
in inhomogeneous winds is prerequisite for the interpretation of O-star
spectra and the determination of mass-loss rates.

\begin{acknowledgements}
Some of the data presented in this paper were obtained from the
Mikulski Archive for Space Telescopes (MAST). STScI is operated by the
Association of Universities for Research in Astronomy, Inc., under NASA
contract NAS5-26555. Support for MAST for non-HST data is provided by
the NASA Office of Space Science via grant NNX09AF08G and by other
grants and contracts. 
\\
The authors would like to thank Dr. Petr \v{S}koda for securing a
spectrum of HD\,192639 in the {\Halpha} region, and to night assistants
Mr. Jan Sloup and Mrs. Lenka Kotkov\'a for their help with obtaining
spectra used in this paper.
\\
This work was supported by grant GA \v{C}R P209/11/1198. B\v S thanks
Ministry of Education and Science of Republic of Serbia who
supported this work through the project  176002 ``Influence of
collisions on astrophysical plasma spectra''.  AA acknowledges
financial support from the research project SF0060030s08 of the
Estonian Ministry of Education and Research. LMO acknowledges support
from DLR grant  50 OR 1302. AFT thanks financial support from the 
Agencia de Promoci\'on Cient\'ifica y Tecnol\'ogica (Pr\'estamo BID 
PICT 2011/0885), PIP 0300 CONICET, and the Programa de Incentivos 
G11/109 of the Universidad Nacional de La Plata, Argentina. Financial 
support from International Cooperation of the Czech Republic 
(M\v SMT, 7AMB12AR021) and Argentina (Mincyt-Meys, ARC/11/10) is 
acknowledged. The Astronomical Institute Ond\v rejov is
supported by the project RVO:67985815. 

\end{acknowledgements}

\bibliographystyle{aa}
\bibliography{Surlan_at_al_astro-ph}

\Online

\let\cleardoublepage\clearpage

\appendix

\onecolumn

\section{Spectral fits}

\begin{figure*}[!b]
\begin{center}
\includegraphics[width=0.86\textwidth]{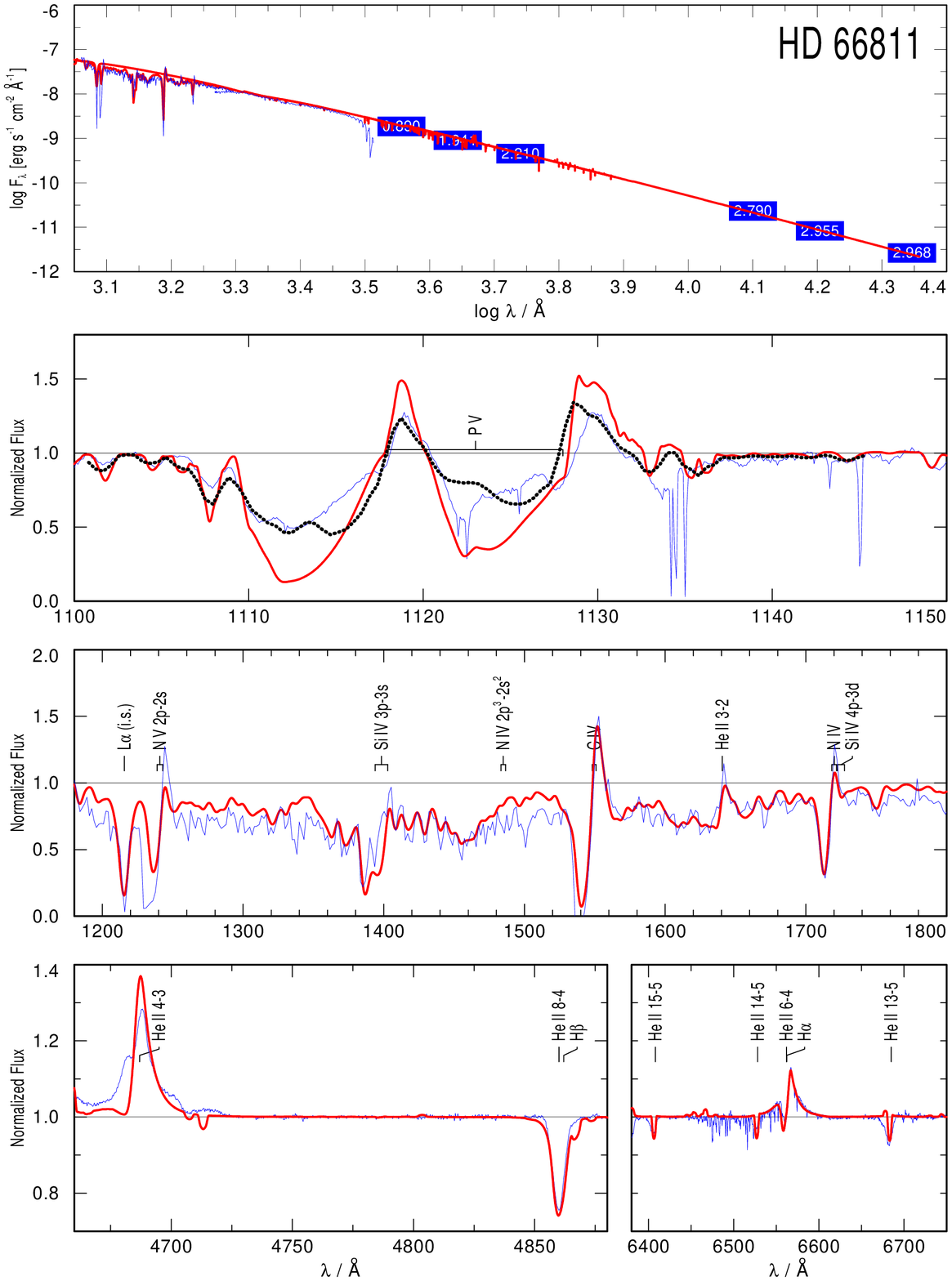}
\end{center}
\caption{Best fit from {\POWR} modeling (thick red-solid lines) to the
observed HD\,66811 spectra (thin blue-solid lines) together with the
calculated {\ionp} line profile from \mbox{3-D} Monte-Carlo code
(black-dotted line). Blue labels with numbers in the upper panels are
UBVJHK magnitudes.}
\label{figzetpup}
\end{figure*}

\begin{figure*}
\begin{center}
\includegraphics[width=0.95\textwidth]{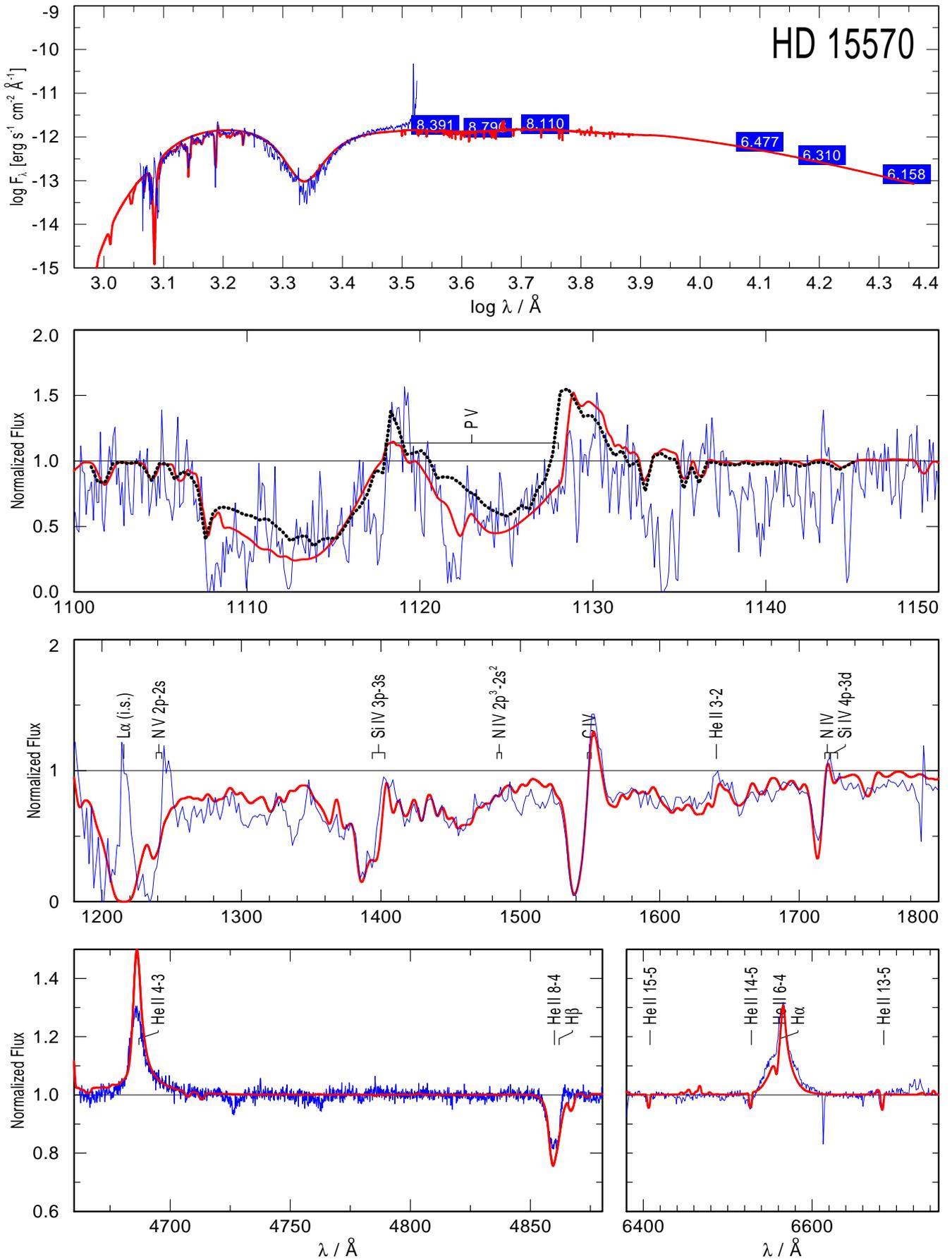}
\end{center}
\caption{The same as Fig.\,\ref{figzetpup}, but for HD\,15570.}
\label{fighd15570}
\end{figure*}

\begin{figure*}
\begin{center}
\includegraphics[width=0.95\textwidth]{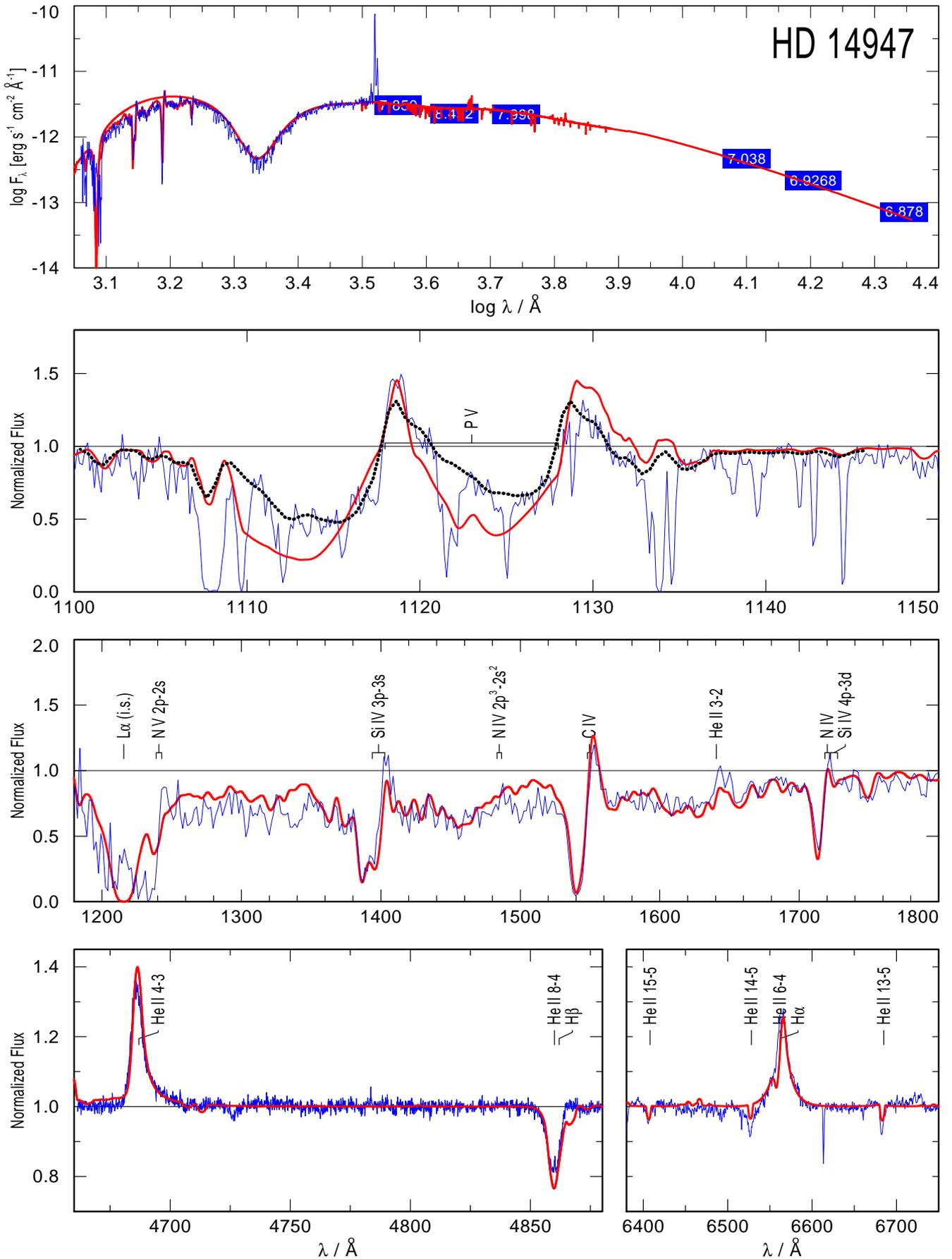}
\caption{The same as Fig.\,\ref{figzetpup}, but for HD\,14947.}
\label{fighd14947}
\end{center}
\end{figure*}

\begin{figure*}
\begin{center}
\includegraphics[width=0.95\textwidth]{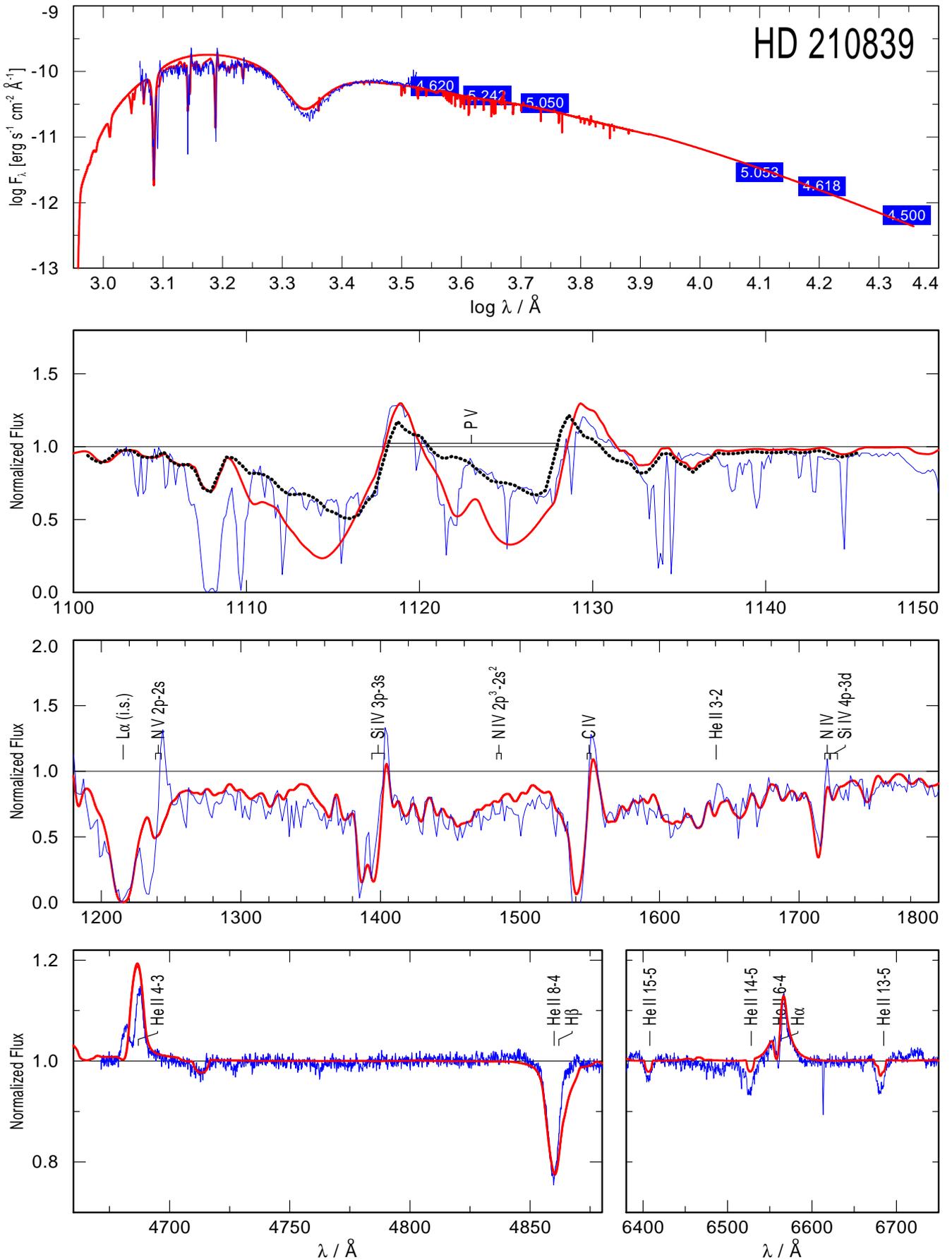}
\caption{The same as Fig.\,\ref{figzetpup}, but for HD\,210839.}
\label{fighd210839}
\end{center}
\end{figure*}

\begin{figure*}
\begin{center}
\includegraphics[width=0.95\textwidth]{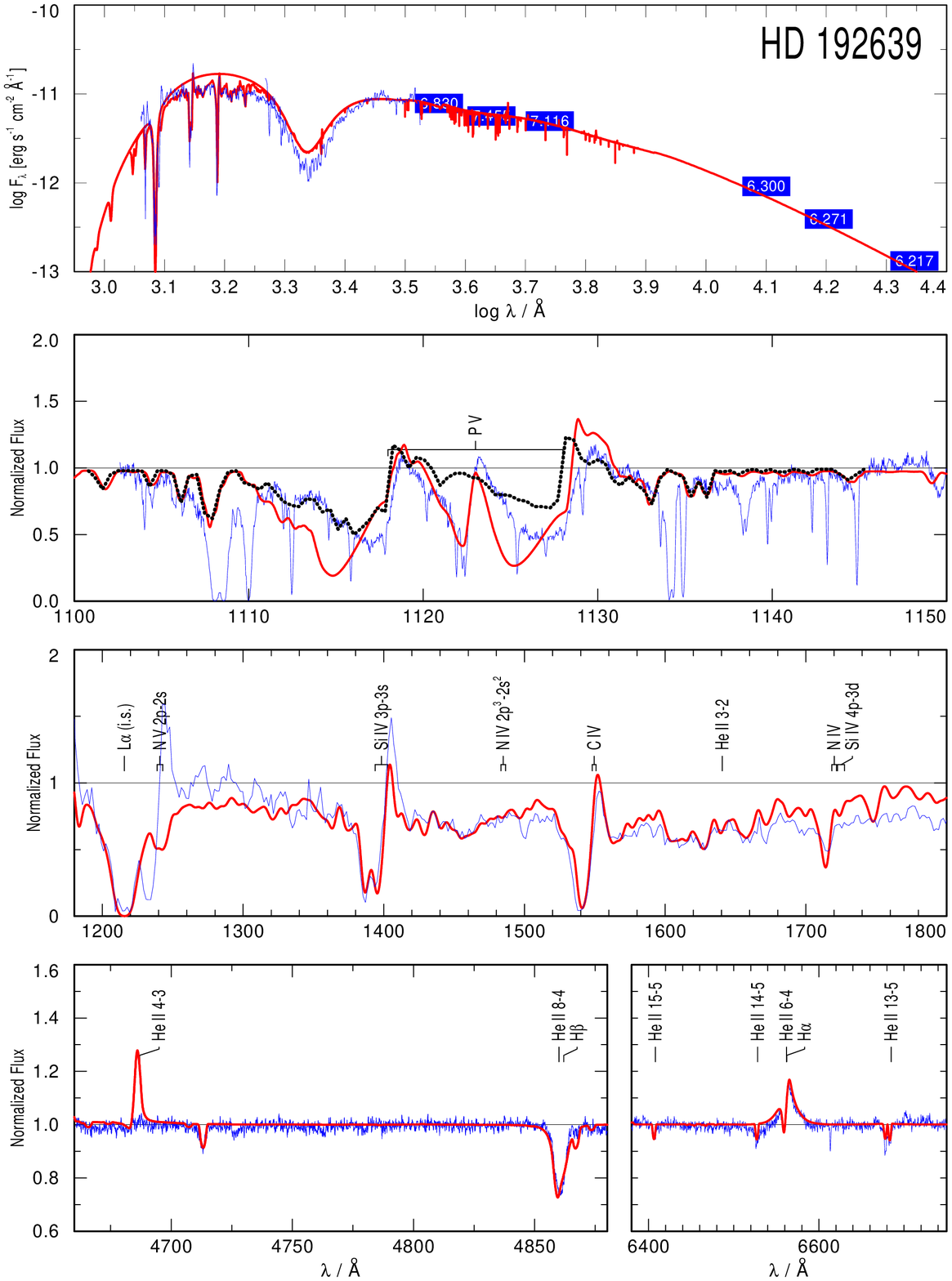}
\caption{The same as Fig.\,\ref{figzetpup}, but for HD\,192639.}
\label{fighd192639}
\end{center}
\end{figure*}

\end{document}